\documentclass[twocolumn]{aastex63} 
\usepackage{epsfig}
\usepackage{graphicx}
\usepackage{float}
\usepackage{amsmath}
\usepackage{color}
\usepackage{amssymb}
\usepackage{amsfonts}
\usepackage{units} 
\usepackage{bm}
\usepackage[mathscr]{euscript}

\def\be{\begin{equation}}
\def\ee{\end{equation}}

\def\bes{\begin{align}}
\def\ees{\end{align}}

\newcommand{\fraction}[3]{\left(\frac{#1}{#2}\right)^{#3}}

\shorttitle{Narrow Spectra of FRBs}
\shortauthors{Yang} 

\begin{document}

\title{Implications of Narrow Spectra of Fast Radio Bursts}

\correspondingauthor{Yuan-Pei Yang} 
\email{ypyang@ynu.edu.cn}

\author[0000-0001-6374-8313]{Yuan-Pei Yang}
\affiliation{South-Western Institute for Astronomy Research, Yunnan University, Kunming, Yunnan 650504, China}
\affiliation{Purple Mountain Observatory, Chinese Academy of Sciences, Nanjing, Jiangsu 210023, China}

\begin{abstract}

Fast radio bursts (FRBs) are millisecond-duration radio transients with extremely high brightness temperatures at cosmological distances, and the physical origin and the radiation mechanism of FRBs are still unknown. The observed spectral bandwidth of some FRBs appeared narrow compared with their peak frequencies, which could be used to constrain the radiation mechanism and the astrophysical environment of FRBs. In this work, we investigate some possible physical origins of the narrow spectra from the perspectives of intrinsic radiation mechanisms, coherent processes, radiative transfers, and interference processes. We find that: (1) If the observed narrow spectra of FRBs are attributed to the intrinsic radiation mechanism by a single charged particle, the particle's deflection angle should be much smaller than the radiation beaming angle. (2) Coherent process can make cause narrow spectra. For the bunching mechanism, the narrow spectra might arise from the radiating bunches with a quasi-periodic distribution. For the maser mechanism, the negative absorption process can naturally cause a narrow spectrum. (3) Most absorption and scattering processes do not significantly change the observed spectra based on the current observation of some FRB repeaters. (4) Scintillation and plasma lensing in the FRB source environment can modulate the spectra, leading to narrow spectra and the burst-to-burst variation of spectra. A planet-like object can generate spectral modulation via gravitational lensing at the GHz band, but the observed burst-to-burst variation of the spectra does not support this scenario. 

\end{abstract} 


\section{Introduction}
Fast radio bursts (FRBs) are millisecond-duration radio bursts with extremely high brightness temperatures of $T_B\sim10^{35}~{\rm K}$, which suggests that their radiation mechanisms must be coherent. Some coherent emission mechanisms have been invoked to interpret the emissions of FRBs, including coherent radiation by charged bunches \citep{Katz14,Katz18,Kumar17,Yang18,Yang23,Lu20,Cooper21,Wang22b,Kumar22b,Qu23}, maser by hydrodynamic instabilities or kinetic instabilities \citep{Lyubarsky21,Beloborodov17,Waxman17,Metzger19}, coherent plasma radiation \citep{Yang21,Mahlmann22}, etc. However, there is no smoking gun to identify the radiation mechanism of FRBs so far.
In addition to the radiation mechanism, the physical origin of FRBs also remains an unsolved puzzle due to the diversity of FRBs \citep[see the recent review of][]{Zhang22b}.
FRB 200428 was detected to be associated with a Galactic magnetar SGR J1935+2154 \citep{Bochenek20,CHIME20,Mereghetti20,Li20,Ridnaia20,Tavani20}, implying that at least some FRBs originate from the magnetars born from the core collapses of massive stars. However, the association between FRB 20200120E and its host globular cluster with an extremely old age challenges the core-collapse magnetar formation \citep{Bhardwaj21,Kirsten22}, which means that it is more likely produced by an old object or a system associated with a compact binary merger \citep{Wang16,Zhang20c,Kremer21,Lu22}.

Up to the present, hundreds of FRB sources have been detected, and dozens of them show repeating behaviors \citep[e.g.,][]{CHIME21}. The increasing number of detected FRBs starts to shed light on the diversity among the phenomena, and the properties of the observed spectra provide important information about the radiation mechanism of FRBs. 
The first CHIME/FRB catalog identified four observed archetypes of burst morphology \citep{Pleunis21}, including simple broadband, simple narrow band, temporally complex, and downward drifting.
Meanwhile, the bursts from FRB repeaters have a larger pulse duration, narrower bandwidth, and lower brightness temperature than those of the one-off FRBs, which might be due to a beaming, propagation effect, or intrinsic populations.
\citet{Law17} made the first simultaneous detection of FRB 20121102A using multiple telescopes and found that its burst spectra could not be well modeled by a power law and more like a Gaussian shape characterized by a $\sim500~{\rm MHz}$ envelope.
\citet{Zhou22} recently reported over 600 bursts from the repeating FRB 20201124A during an active episode and found that the sub-bursts of FRB 20201124A show narrow spectra with a center frequency of $1.09~{\rm GHz}$ and a characteristic bandwidth of $\sim 277{\rm MHz}$. 
FRB 20220912A also has many bursts with narrow spectral bandwidth \citep{Zhang23}. For the bursts with their spectra within the L band of the Five-hundred-meter Aperture Spherical radio Telescope (FAST), the relative spectral bandwidth of the radio bursts was found to be distributed near $\Delta\nu/\nu_0\sim(0.1-0.2)$.
Some FRBs show more extremely narrow bandwidth. One burst of FRB 20190711A has a central frequency of $1.4~{\rm GHz}$ and a full-width-at-half-maximum (FWHM) bandwidth of just $65~{\rm MHz}$, and no evidence of any emission in the remaining part of the $3.3~{\rm GHz}$ band of the Ultra-wideband Low (UWL) receiver system of the Parkes radio telescope \citep{Kumar21b}, which means that the relative spectral bandwidth is only $\Delta\nu/\nu_0\sim0.05$.

In this work, we will discuss the possible physical origins of the observed narrow spectra of FRBs from the perspectives of intrinsic radiation mechanisms, coherent processes, radiative transfers, and interference processes. The paper is organized as follows. In Section \ref{sec_narrow}, we discuss the spectral bandwidth distribution and the possible physical processes affecting the FRB spectra. 
In Section \ref{mechanism}, we generally analyze the radiation features of the intrinsic radiation mechanisms by a single charged particle, including the radiation mechanisms with the particle's deflection angle larger than the radiation beaming angle in Section \ref{mechanism1} and the opposite scenario in Section \ref{mechanism2}, and the possible astrophysical scenarios are discussed in the Section \ref{picture}. 
In Section \ref{Section_Coherence}, we discuss how the coherent processes change the radiation spectra, including the bunching mechanism in Section \ref{Coherence_B} and the maser mechanism in Section \ref{Coherence_M}.
The radiative transfers (including absorption and scattering processes) are discussed in Section \ref{transfer}, and some interference processes (scintillation, gravitational lensing, and plasma lensing) at a large-scale region are discussed in Section \ref{propagation}.
The results are summarized and discussed in Section \ref{summary}.
The convention $Q_x=Q/10^x$ is adopted in cgs units unless otherwise specified.
Some detailed calculations are presented in the Appendices.

\section{Narrow spectra of FRBs: observation and physical origin}\label{sec_narrow}

\subsection{Spectral bandwidth distribution of FRB repeaters}\label{specband}

Some FRBs, e.g., FRB 20190711A, FRB 20201124A, and FRB 20220912A, appear extremely narrow spectra in the bandwidths of telescopes \citep[e.g.,][]{Kumar21b,Zhou22,Zhang23}, implying that the spectra of at least some FRBs are intrinsically narrow. In this section, we will discuss the implication of the observed narrow bandwidths of FRBs, and emphasize that the observed relative spectral bandwidth mainly depends on the intrinsic spectral shape, the bandwidth definition (e.g., full width at half maximum, full width at tenth maximum, etc.), and the telescope's bandwidth.
We first consider that the intrinsic spectra of radio bursts from an FRB source have a general form described by a broken power-law distribution,
\begin{align}
F_\nu=F_{\nu,0}\left\{
\begin{aligned}
&\fraction{\nu}{\nu_0}{\alpha_l},&&\text{for}~\nu\leqslant\nu_0,\\
&\fraction{\nu}{\nu_0}{-\alpha_h},&&\text{for}~\nu>\nu_0,
\end{aligned}
\right.
\end{align}
where $F_{\nu,0}$ corresponds to the maximum flux at the peak frequency $\nu_0$. We should notice that $\alpha_l>0$ and $\alpha_h>0$ are assumed here, considering that the flux $F_\nu$ vanishes at $\nu\rightarrow0~{\rm or}~\infty$.
In literature, one usually defines the spectral bandwidth via the full width at a fraction of maximum ($F_{\nu,0}/N_{\rm FW}$), e.g., the full width at half maximum (FWHM) with $N_{\rm FW}=2$ \citep[e.g.,][]{Kumar21b,Zhang23} or via the full width at tenth maximum (FWTM) with $N_{\rm FW}=10$ \citep[e.g.,][]{Pleunis21}.
Thus, the relative spectral bandwidth of a radio burst is
\begin{align}
\frac{\Delta\nu}{\nu_0}\simeq N_{\rm FW}^{{\frac{1}{\alpha_h}}}-N_{\rm FW}^{{-\frac{1}{\alpha_l}}}.\label{SN2}
\end{align}
We can see that the relative spectral bandwidth $\Delta\nu/\nu_0$ depends on the factor of $N_{\rm FW}$ and the intrinsic spectral shape that is described by the above two power-law indexes. 
\emph{Since such a defined (FWHM or FWTM) relative spectral bandwidth only depends on the intrinsic spectral shape, the $\Delta\nu/\nu_0$ distribution of an FRB repeater should be narrow if the intrinsic spectral shape keeps unchanged.} 

\begin{table*}
\begin{center}
    \caption{Summary of the low-frequency spectral index $\alpha_l$ for various radiation mechanisms}
    \resizebox{\linewidth}{!}{
    \begin{tabular}{ccc}
    \hline\hline
    Radiation mechanisms     &Low-frequency spectral index $\alpha_l$   & References\\
    \hline\hline
    Curvature radiation by a single charged particle &2/3    &\citet{Jackson98,Yang18}\\
    \hline
    Curvature radiation by a bunch-cavity (or electron-positron) pair &8/3    &\citet{Yang20,Yang23}\\
    \hline
    Curvature radiation by fluctuating bunches &0    &\citet{Yang23}\\
    \hline
    Synchrotron radiation by particles with a random pitch-angle distribution &1/3    &\citet{Jackson98,Rybicki86}\\
    \hline
    Synchrotron radiation by particles with a narrow pitch-angle distribution &2/3    &\citet{Yang18b}\\
    \hline
    Synchrotron self-absorption &5/2    &\citet{Rybicki86}\\
    \hline
    Jitter radiation &1    &\citet{Medvedev00,Dermer09}\\
    \hline
    Blackbody radiation &2    &\citet{Rybicki86}\\
    \hline
    Bremsstrahlung radiation &0    &\citet{Rybicki86}\\
    \hline
    Inverse Compton scattering & Depend on incident photon spectrum &\citet{Rybicki86,Zhang22}\\
    \hline\hline
    \end{tabular}}\label{table}
\end{center}
\end{table*}

Most radiation mechanisms involved in various astrophysical scenarios have a low-frequency spectral index of $\alpha_l<3$, see Table \ref{table}. 
For example, the synchrotron self-absorption has a low-frequency spectral index of $\alpha=5/2$ \citep{Rybicki86} and the curvature radiation by a bunch-cavity (or electron-positron) pair has a low-frequency spectral index of $\alpha=8/3$ \citep{Yang20,Yang23}.
Thus, for these radiation mechanisms with $\alpha_l<3$, according to Eq.(\ref{SN2}), the relative spectral bandwidth for FWHM with $N_{\rm FW}=2$ should satisfy
\begin{align}
\frac{\Delta\nu}{\nu_0}>0.2.
\end{align}

For example, we assume that the intrinsic spectrum is due to the curvature radiation, then one approximately has $\alpha_l\simeq2/3$ and $\alpha_h\rightarrow\infty$ \citep[][]{Yang18,Yang23}. According to Eq.(\ref{SN2}), the relative spectral bandwidth for FWHM is $\Delta\nu/\nu_0\simeq0.65$.
In particular, for FRB 20190711A with an extremely narrow FWHM spectral bandwidth of $\Delta\nu/\nu_0\sim0.05$ \citep{Kumar21b}, one has $\min(\alpha_l,\alpha_h)\gtrsim 14$, implying an extremely narrow intrinsic spectrum that should involve some special mechanisms (that might be attributed intrinsic radiation mechanisms, coherent processes, radiative transfers, or interference processes), see the following discussions.

In reality, the bandwidth $\Delta\nu_t$ of a radio telescope is usually narrow compared with the telescope's central frequency $\nu_{0,t}$, $\Delta\nu_t\ll\nu_{0,t}$. For example, the L band of FAST is from 1 GHz to 1.5 GHz, i.e., $\nu_{0,t}=1.25~{\rm GHz}$ and $\Delta\nu_t=0.5~{\rm GHz}$.
Due to the limited telescope's bandwidth, many observed spectra of FRBs are often incomplete. 
If a radio burst is observable for a certain telescope, its emission must be within the telescope's bandwidth, leading to the following conditions:
\begin{align}
\nu_0+\frac{\Delta\nu}{2}>\nu_{0,t}-\frac{\Delta\nu_t}{2}~~~{\rm and}~~~\nu_0-\frac{\Delta\nu}{2}<\nu_{0,t}+\frac{\Delta\nu_t}{2}.
\end{align}
The observed central frequency $\nu_{0,\rm obs}$ (not the intrinsic peak frequency) of an observable FRB is in the telescope's bandwidth,
\begin{align}
\nu_{0,\rm obs}\in\left[\nu_{0,t}-\frac{\Delta\nu_t}{2},\nu_{0,t}+\frac{\Delta\nu_t}{2}\right],
\end{align}
although the intrinsic peak frequency $\nu_0$ might be estimated outside the telescope's bandwidth based on the observed spectral shape.
The observed spectral bandwidth of an observable FRB is
\begin{align}
\Delta\nu_{\rm obs}&=\min\left(\nu_0+\frac{\Delta\nu}{2},\nu_{0,t}+\frac{\Delta\nu_t}{2}\right)\nonumber\\
&-\max\left(\nu_0-\frac{\Delta\nu}{2},\nu_{0,t}-\frac{\Delta\nu_t}{2}\right).
\end{align}
Since the bandwidth of a radio telescope is usually narrow, the distribution of the intrinsic peak frequency $\nu_0$ of an FRB repeater could be approximately assumed to be uniform near the telescope's bandwidth, i.e., the distribution function of $\nu_0$ could be approximately described by
\begin{align}
f(\nu_0)\sim {\rm const}.
\end{align}

\begin{figure}
    \centering
    \includegraphics[width = 0.9\linewidth, trim = 0 0 0 0, clip]{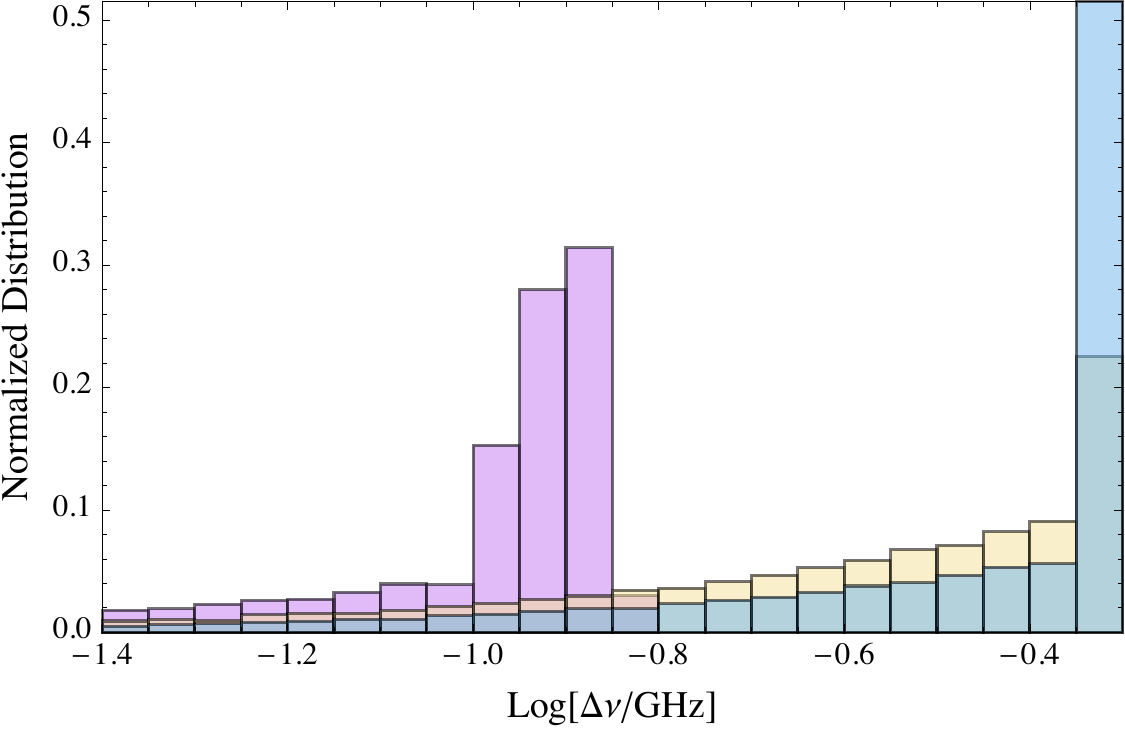}
    \caption{The simulated distribution of the observed spectral bandwidth of radio bursts from an FRB repeater. The purple, yellow, and blue bars correspond to the different relative spectral bandwidths with $\Delta\nu/\nu_0=0.1,0.5~{\rm and}~1$, respectively. We take $\nu_{0,t}=1.25~{\rm GHz}$ and $\Delta\nu_t=0.5~{\rm GHz}$ for the telescope's parameters. The intrinsic peak frequencies $\nu_0$ of the radio bursts are assumed to be uniformly distributed near the telescope's bandwidths, i.e., $f(\nu_0)={\rm const}$.}\label{bandwidth} 
\end{figure}

We make a Monte Carlo simulation to generate the distribution of the observed spectral bandwidth $\Delta\nu_{\rm obs}$ of the radio bursts from an FRB repeater, as shown in Figure \ref{bandwidth}, and take $\nu_{0,t}=1.25~{\rm GHz}$ and $\Delta\nu_t=0.5~{\rm GHz}$ that are consistent with the L band of FAST. For $\Delta\nu/\nu_0=0.1$, the observed spectral bandwidth $\Delta\nu_{\rm obs}$ of most bursts are consistent with the intrinsic ones due to $\Delta\nu<\Delta\nu_t$. For $\Delta\nu/\nu_0=1$, since many bursts have $\Delta\nu>\Delta\nu_t$, the observed spectral bandwidth $\Delta\nu_{\rm obs}$ of most bursts would be constrained by the bandwidth of the telescope, leading to $\Delta\nu_{\rm obs}\sim\Delta\nu_t$. 
Very few bursts have $\Delta\nu_{\rm obs}<\Delta\nu_t$, because only a part of one end (low-frequency end or high-frequency end) of the intrinsic spectral bandwidth $\Delta\nu$ is within the telescope's bandwidth. 
Considering that many radio bursts are incomplete in the frequency domain due to the narrow telescope's bandwidth, the observed spectral bandwidth distribution of radio bursts from an FRB repeater could be used to test whether most of the bursts' spectra of an FRB source are intrinsically narrow.

\subsection{Physical origin of FRB narrow spectra}\label{nspec}

Next, we generally discuss the possible physical origin of the intrinsic narrow spectra of FRBs.
We consider that a finite pulse of electromagnetic wave has the form of $\vec{E}(t)=\vec{E}_\parallel(t)+\vec{E}_\perp(t)$, where $\vec{E}_\parallel(t)$ and $\vec{E}_\perp(t)$ are a pair of orthogonal components of $\vec{E}(t)$. The properties of $\vec{E}(t)$ vary with time and vanishes sufficiently rapidly for $t\rightarrow\pm\infty$, and the power spectrum of $\vec{E}(t)$ satisfies $|E(\omega)|^2=|E_\parallel(\omega)|^2+|E_\perp(\omega)|^2$.
Since the two orthogonal components are independent, let us treat only one of the two components, $\vec{E}_k(t)$ with $k=\parallel,\perp$.
In particular, if an observed spectrum $|E(\omega)|^2$ appears narrow, the main component between $|E_\parallel(\omega)|^2$ and $|E_\perp(\omega)|^2$ must also be narrow. 

\begin{figure}
    \centering
    \includegraphics[width = 0.9\linewidth, trim = 0 0 0 0, clip]{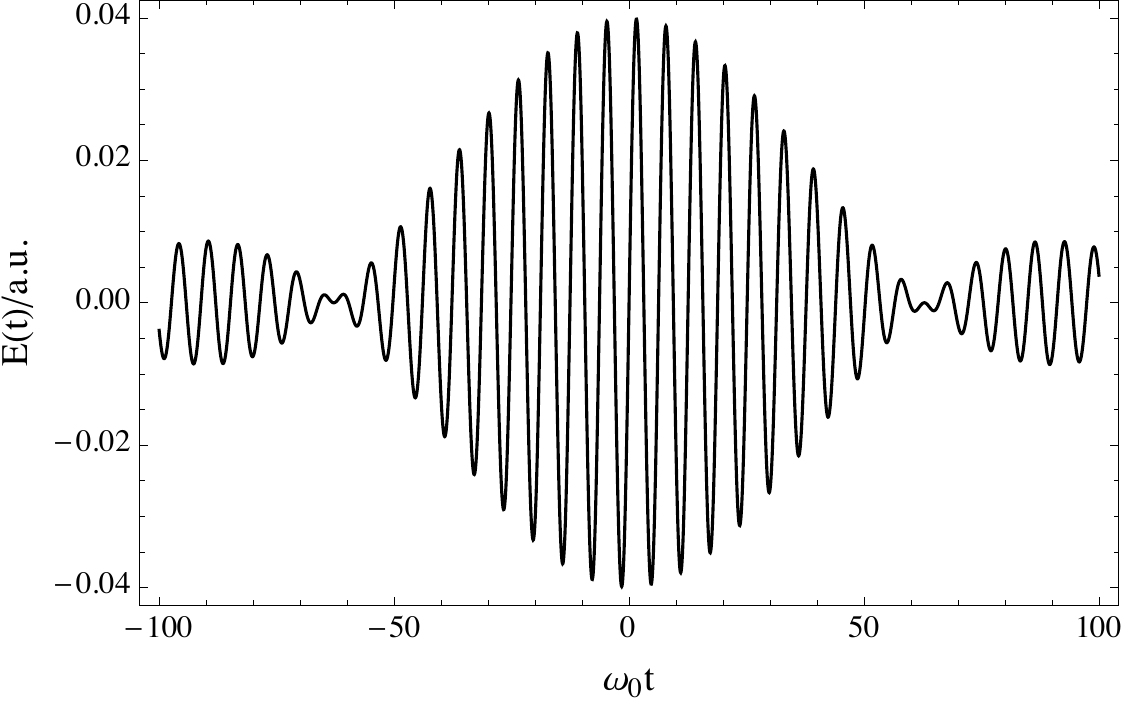}
    \includegraphics[width = 0.9\linewidth, trim = 0 0 0 0, clip]{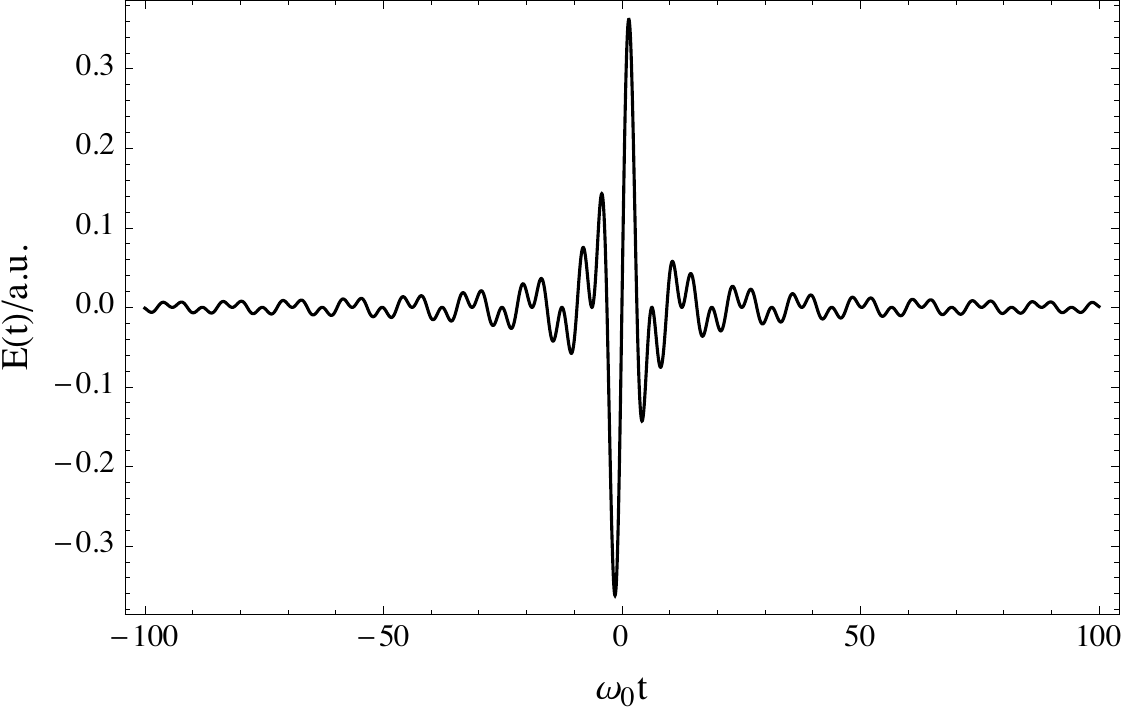}
    \caption{The evolution of electric field component of a pulse of electromagnetic wave with a rectangular power spectrum given by Eq.(\ref{rect}). The top and bottom panels correspond to the scenarios with a narrow spectrum of $\Delta\omega_k/\omega_{0,k}=0.1$ and with a wide spectrum of $\Delta\omega_k/\omega_{0,k}=1$, respectively. The phase argument is taken as $\phi_k=0$ here.}\label{Et}
\end{figure}

Without loss of generality, the spectrum of the main component could be roughly described by a rectangular profile with a central frequency $\omega_{0,k}$ and a spectral bandwidth $\Delta\omega_k$,
\begin{align}
|E_k(\omega)|^2\propto{\rm rect}\left(\frac{\omega-\omega_{0,k}}{\Delta\omega_k}\right),\label{rect}
\end{align}
where ${\rm rect}(x)$ is the rectangular function that is defined as ${\rm rect}(x)=1~\text{for}~|x|\leqslant1/2$ and ${\rm rect}(x)=0~\text{for}~|x|>1/2$. 
Thus, $E_k(\omega)$ as the Fourier transformation of $E_k(t)$ is given by 
\begin{align}
E_k(\omega)\propto{\rm rect}\left(\frac{\omega-\omega_{0,k}}{\Delta\omega_k}\right)e^{i\phi_k},
\end{align}
where $\phi_k$ is a phase argument.
Generally, $\phi_k$ cannot be directly obtained only based on the information of the power spectrum $|E_k(\omega)|^2$, but for a selected appropriate pair of orthogonal components, $E_\parallel(\omega)$ and $E_\perp(\omega)$ might be a pair of real and imaginary numbers, i.e., ${\rm Im}[E_\parallel(\omega)]=E_\parallel(\omega)$ and ${\rm Re}[E_\perp(\omega)]=E_\perp(\omega)$, see the following discussions in Section \ref{mechanism1} and Section \ref{mechanism2}, leading to $\phi_k(\omega)=n\pi/2$ with $n\in\mathbb{Z}$. In this case, one may take $\phi_k={\rm const}$.
According to the properties of the Fourier transform, the corresponding pulse profile is
\begin{align}
E_k(t)\propto\frac{\Delta\omega_k}{\sqrt{2\pi}}{\rm sinc}\frac{\Delta\omega_k t}{2}e^{-i(\omega_{0,k}t-\phi_k)},\label{Ekt}
\end{align}
where ${\rm sinc}(x)\equiv\sin x/x$. 
In Figure \ref{Et}, we plot the pulse profile $E_k(t)$ based on Eq.(\ref{Ekt}). The top panel shows the scenario with a narrow spectrum of $\Delta\omega_k/\omega_{0,k}=0.1$ and the bottom panel shows the scenario with a wide spectrum of $\Delta\omega_k/\omega_{0,k}=1$. 
We can see that\footnote{Notice that although the spectrum by Eq.(\ref{rect}) is assumed to be a rectangular profile here for analytical analysis, this conclusion is widely applicable to other spectral profiles. For example, if the spectrum is described by a Gaussian profile with a peak frequency $\omega_{0,k}$ and a scatter of $\Delta\omega_k$, the corresponding electromagnetic signal would also be periodic and quasi-sinusoid with a frequency of $\omega\sim\omega_{0,k}$ in a short term and have a typical pulse duration of $T\sim4\pi/\Delta\omega_k$ in a long term, which could be easily tested via the numerical calculation of the corresponding Fourier transform. } 
a narrow spectrum with $\Delta\omega_k/\omega_{0,k}\ll1$ implies that \emph{the electromagnetic signal $E_k(t)$ should be periodic and quasi-sinusoid with an oscillating frequency of $\omega\sim\omega_{0,k}$ in a short term} and have a typical pulse duration of $T\sim4\pi/\Delta\omega_k$ in a long term. 
This conclusion is natural because a narrow spectrum means that the radiation should be quasi-monochromatic, leading to a quasi-sinusoid periodic waveform of $E_k(t)$.
In particular, if the periodic quasi-sinusoid signal $E_k(t)$ is produced by the intrinsic radiation mechanism of a single charged particle, and the particle's acceleration is required to be periodic during its radiation beam pointing to the observer, see the discussion in Section \ref{mechanism2} and Appendix \ref{mechanism2a}.

\begin{figure}
    \centering
    \includegraphics[width = 0.8\linewidth, trim = 100 50 100 50, clip]{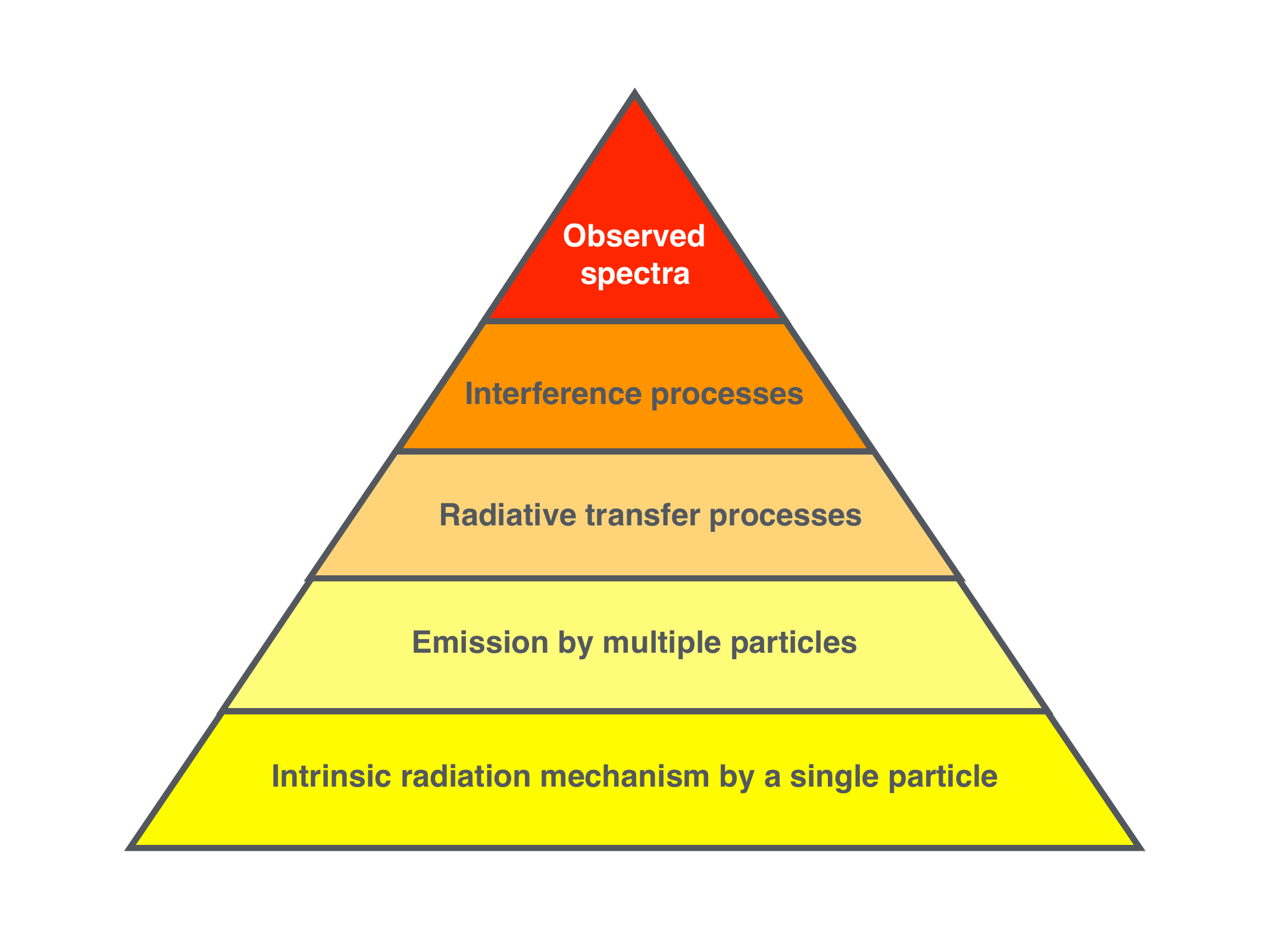}
    \caption{The schematic pyramid illustrates the physical processes affecting the observed spectral features.}\label{pyramid}
\end{figure}

In general, the observed spectral features of an astrophysical phenomenon mainly depend on the following physical processes from the fundamental level to the highest level, see Figure \ref{pyramid}: 

(1) The most fundamental level is the intrinsic radiation mechanism by a single charged particle that determines the initial radiation of each particle in the emission region. 
Since the radiating particles in one emission region usually have different energies, in most astrophysical phenomena, the information on the intrinsic radiation mechanism is directly reflected by the spectral shape at the lowest band that is contributed by the particles with the lowest energies.

(2) The second level is the total emission by all radiating particles in the emission region, which determines the emission coefficient of the emission system. There are two scenarios in this level: incoherent radiation and coherent radiation. For incoherent radiation, the radiation spectral power is just a simple summation of the spectral power of each particle. In this case, due to a wide distribution of the parameters of multiple particles, the relative spectral bandwidth $\Delta\nu/\nu_0$ should be at least wider than that of the intrinsic radiation mechanism by a single particle. 
However, for coherent radiation, because of the superposition of electromagnetic waves at certain frequencies, the total spectral shape not only depends on the intrinsic radiation mechanism by a single particle but also depend on the specific coherent process in the emission region.

(3) The third level corresponds to the radiative transfer processes, which involves absorption\footnote{If the particles have inverted population, the absorption coefficient would become negative, in which case, rather than decrease along ray, the intensity actually increases, so-called as ``maser''. However, since the maser mechanism is also one of the main coherent radiation mechanisms, we particularly classify it as the second level in this paper.} and scattering. Due to these radiative transfer processes, the incident emission by radiating particles would be changed by absorption or scattering in certain regions. In particular, the absorption processes (e.g., free-free absorption, plasma absorption, synchrotron self-absorption, etc.) usually make the spectra harder, leading to a significant cutoff at the low-frequency band. The frequency-independent electron scattering processes simultaneously suppress the radiation power in all bands, thus, the spectral shape keeps unchanged.

(4) The highest level is the interference processes that change the radiation spectra via the wave interference at a large-scale region, including scintillation, gravitational lensing, plasma lensing, etc. These interference processes might occur in the FRB environments (e.g., circumburst medium, companion wind in a binary system, etc.) or at some regions far away from the FRB sources (e.g., interstellar medium, intergalactic medium, etc.).

In the following sections, we will analyze the above processes in detail and discuss the corresponding implications for the FRB observation.

\section{Radiation spectra by a single charged particle}\label{mechanism}

\begin{figure}
    \centering
    \includegraphics[width = 0.9\linewidth, trim = 100 150 100 150, clip]{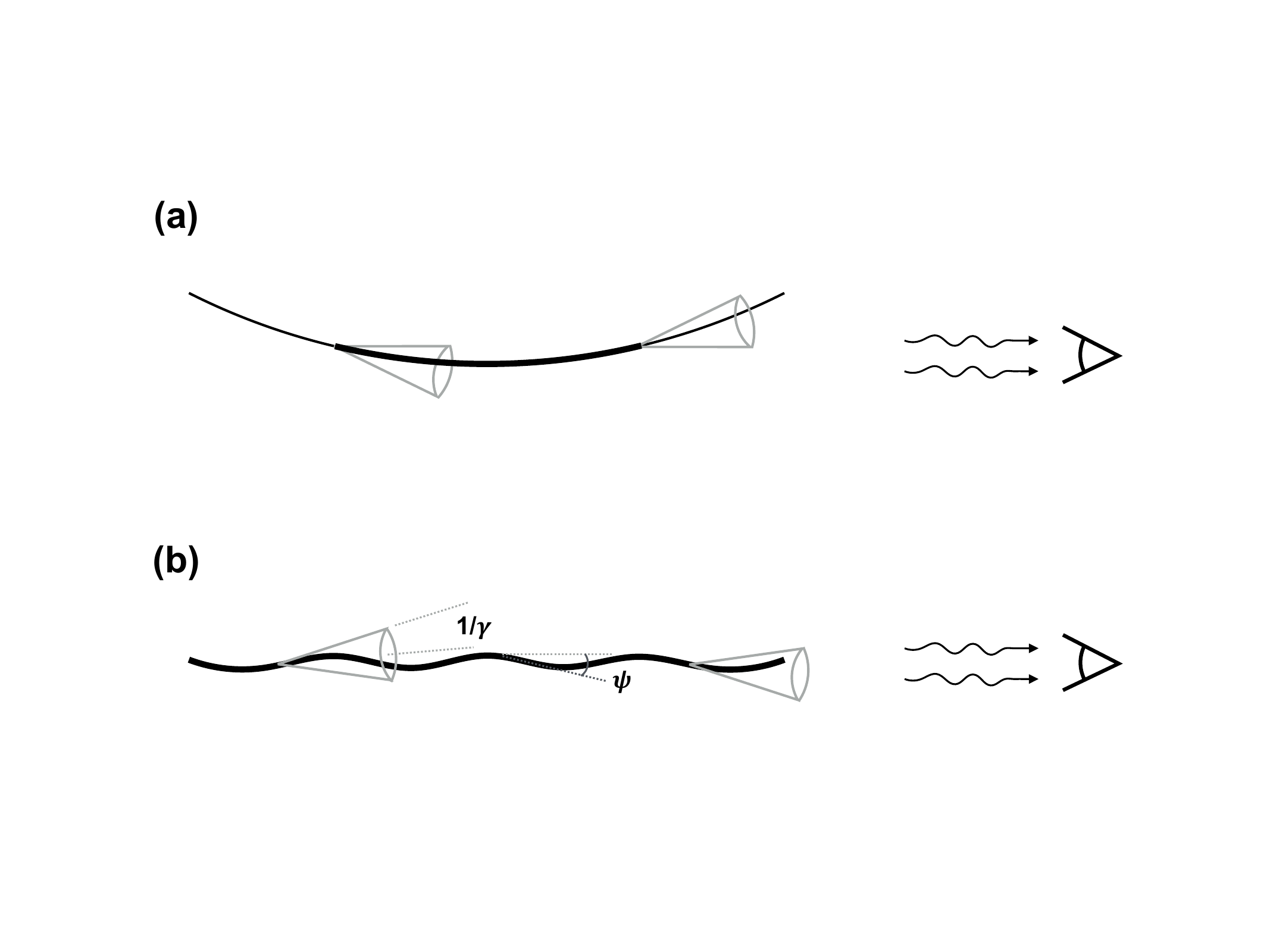}
    \caption{Emission from various points along the trajectory of a relativistic particle. Panel (a) $\gamma\psi\gg1$: the emission from some parts (bold portions) of the trajectory is observable. Panel (b) $\gamma\psi\ll1$: the emission from the entire trajectory is observable.}\label{emission_trajectory} 
\end{figure}

We first discuss the spectral features of the intrinsic radiation mechanisms by a single charged particle, as the fundamental level pointed out above.
The radiation of a single charged particle with Lorentz factor $\gamma$ undergoing arbitrary accelerated motion is a coherent superposition with the contributions from the accelerations parallel to and perpendicular \emph{to the particle's velocity}. For comparable parallel and perpendicular forces, the radiation from the parallel component is of order $1/\gamma^2$ compared to that from the perpendicular component. Thus, one usually neglects the parallel acceleration\footnote{However, we should notice that the parallel acceleration could be dominant under the scattering process. For example, if the incident electromagnetic wave is linear polarized and weak (the Lorentz force by the magnetic field component is much weaker than the electric field force), the charged particle would be linearly accelerated by the oscillating electric field (the scenario for strong wave could be seen in \citet{Yang20b}). Besides, under the magnetosphere of a neutron star, even if the incident wave is circularly polarized or strong, the charged particle can only oscillate along the field lines due to the existence of a strong background magnetic field and emit the scattering wave due to the parallel acceleration \citep{Beloborodov22,Zhang22,Qu23b}.}.
The radiation spectrum of the perpendicular component depends on the relation between the particle's deflection angle $\psi$ and the radiation beaming angle $\sim1/\gamma$ \citep{Landau75}, as shown in Figure \ref{emission_trajectory}. The particle's deflection angle $\psi$ is determined as follows. The particle's momentum is $p\sim\gamma m_ec$, and the change in the perpendicular momentum due to a transverse force $F_\perp$ is $p_\perp\sim F_\perp\Delta t_{\rm acc}$ ($\Delta t_{\rm acc}$ is the time during which the particle acceleration changes significantly). Thus, the particle's deflection angle is
\begin{align}
\psi\sim\frac{p_\perp}{p}\sim\frac{F_\perp\Delta t_{\rm acc}}{\gamma m_e c},
~~~\text{leading to}~~~
\gamma\psi\sim\frac{F_\perp\Delta t_{\rm acc}}{m_e c}.
\end{align}
If $\gamma\psi\gg1$, i.e., the particle's deflection angle is much larger than the radiation beaming angle, the observer will see radiation from a short segment of the electron's trajectory that is nearly parallel to the line of sight, as shown in the panel (a) of Figure \ref{emission_trajectory}, which corresponds to the scenarios of curvature radiation \citep[e.g.,][]{Jackson98,Yang18,Yang23}, traditional (large-pitch-angle) synchrotron radiation \citep[e.g.,][]{Ginzburg69,Jackson98,Rybicki86}, etc.
If $\gamma\psi\ll1$, i.e., the particle's deflection angle is much smaller than the radiation beaming angle, the particle's entire trajectory would be seen by the observer, as shown in the panel (b) of Figure \ref{emission_trajectory}, which corresponds to the small-pitch-angle synchrotron radiation \citep[e.g.,][]{Epstein73}, jitter radiation \citep[e.g.,][]{Medvedev00}, etc.
In the following discussion, we will discuss the case of $\gamma\psi\gg1$ in Section \ref{mechanism1} and the case of $\gamma\psi\ll1$ in Section \ref{mechanism2}.

\subsection{Deflection angle larger than radiation beaming angle}\label{mechanism1}

The radiation for $\gamma\psi\gg1$ is equivalent to the radiation by the particle moving instantaneously at constant speed on an appropriate circular path \citep{Jackson98}, as shown in the panel (a) of Figure \ref{emission_trajectory}. 
We consider that the acceleration curvature radius is $\rho$, the angle between the line of sight and the trajectory plane is $\theta$, and the radiation angular frequency is $\omega$.
The radiation energy per unit frequency interval per unit solid angle is \citep{Jackson98}
\begin{align}
\mathcal{E}_\omega=\frac{3e^2}{4\pi^2c}\gamma^2\hat\omega^2(1+\gamma^2\theta^2)^2\left[K_{2/3}^2(\xi)+\frac{1}{1/\gamma^2\theta^2+1}K_{1/3}^2(\xi)\right].\label{CRspec0}
\end{align}
where $\xi=(\hat\omega/2)(1+\gamma^2\theta^2)^{3/2}$, $\hat\omega=\omega/\omega_c$, and $\omega_c=3\gamma^3c/2\rho$ is the typical radiation frequency.

\emph{The radiation spectrum by a single radiating particle with $\gamma\psi\gg1$ is intrinsically wide} \citep{Jackson98,Yang18}, $\Delta\omega/\omega_0\sim1$, as discuss in Appendix \ref{mechanism1a} in detail, and it satisfies the power-law distribution with $\mathcal{E}_\omega\propto\hat\omega^{2/3}$ at the low frequency and appears an exponential decay at the high frequency. Therefore, the observed narrow spectrum of FRBs can not be attributed to the intrinsic radiation mechanism by a single charged particle with $\gamma\psi\gg1$ (also see \citet{Katz18} for a detailed discussion).

The polarization properties for the scenario of $\gamma\psi\gg1$ are discussed in Appendix \ref{mechanism1a}. For the radiation by a single charged particle, the intrinsic linear/circular polarization degree mainly depends on the angle between the viewing direction and the trajectory plane. The larger the viewing angle, the lower (higher) the linear (circular) polarization degree. Besides, the higher the observed frequency, the lower (higher) the linear (circular) polarization degree. 
Therefore, the high circular polarization degree should be attributed to the off-beam observation.
If there are multiple radiating particles uniformly distributed in a fan beam, the cumulative distributions of the linear and circular polarization degrees would depend on the telescope’s sensitivity, the particles' beaming angle and the observed frequency.
Important conclusions for the cumulative polarization distributions include: (1) The higher the telescope’s sensitivity, the lower the number fraction between the linearly and circularly polarized bursts. The reason is the bursts at larger viewing angles have higher circular polarization degrees and lower fluxes.
(2) The larger the particles' beaming angle, the higher the number fraction between the linearly and circularly polarized bursts. If the viewing angle is much larger than $1/\gamma$, most bursts would have high linear polarization.
(3) The higher the observed frequency, the higher the number fraction between the linearly and circularly polarized bursts. The reason is that the threshold viewing angle is significantly suppressed at the high frequency, leading to a larger relative number of bursts within the particle beaming angle.

\subsection{Deflection angle smaller than radiation beaming angle}\label{mechanism2}

In the scenario with $\gamma\psi\ll1$, the particle with a charge $q$ moves along the line of sight with an almost constant velocity $\vec{\beta}$ but with a varying acceleration $\dot{\vec{\beta}}$ as shown in the panel (b) of Figure \ref{emission_trajectory}, which is called a ``wiggler'' in the laboratory.
The radiation energy per unit frequency interval per unit solid angle at the line-of-sight direction $\vec{n}$ could be written as \citep[][also see Appendix \ref{mechanism_appendix}]{Landau75}
\begin{align}
\mathcal{E}_\omega
=\frac{q^2}{4\pi^2c}\left(\frac{\omega}{\tilde\omega}\right)^4\left|\vec{n}\times\left[(\vec{n}-\vec{\beta})\times\dot{\vec{\beta}}_{\tilde\omega}\right]\right|^2
\end{align}
with
\begin{align}
\dot{\vec{\beta}}_{\tilde\omega}\equiv\int_{-\infty}^{\infty}\dot{\vec{\beta}}e^{i\tilde\omega t'}dt'~~~\text{and}~~~\tilde\omega\equiv(1-\vec{n}\cdot\vec{\beta})\omega.
\end{align}
where $t'$ is the retarded time.
In the ultrarelativistic case, the longitudinal acceleration is smaller than the transverse acceleration, $\dot{\vec{\beta_\parallel}}/\dot{\vec{\beta_\perp}}\sim1/\gamma^2\ll1$. Thus, $\dot{\vec{\beta}}$ and $\vec{\beta}$ are approximately perpendicular to each other, $\dot{\vec{\beta}}\perp\vec{\beta}$.
Considering that Fourier component $\dot{\vec{\beta}}_{\tilde\omega}$ is significantly different from zero only if $1/\tilde\omega$ is of the same order as the time $\Delta t_{\rm acc}$ during which the particle acceleration changes significantly, the typical frequency of the radiation is estimated to be \citep{Landau75}: 
\begin{align}
\omega\sim(1-\beta)^{-1}\Delta t_{\rm acc}^{-1}\sim2\gamma^2\Delta t_{\rm acc}^{-1}.\label{freq10}
\end{align}
For example, if the particle's acceleration is due to the Lorentz force by the magnetic field $B$, one has $\Delta t_{\rm acc}^{-1}\sim\omega_B=eB/\gamma m_e c$, where $\omega_B$ is the cyclotron frequency, leading to the typical radiation frequency to be $\omega\sim\gamma eB/m_e c$.
 
In many astrophysical scenarios, the perpendicular acceleration of a charged particle is usually attributed to the Lorentz force by magnetic fields. Such a scenario corresponds to the small-pitch-angle synchrotron radiation, also see Appendix \ref{mechanism2a} for a detailed discussion. According to Eq.(\ref{P_omega}), the radiation power per unit frequency interval per unit solid angle is
\begin{align}
\mathcal{P}_\omega=\frac{e^2\gamma^5\psi^2\omega_0}{4\pi c}\bar\omega^4\left(1-\bar\omega+\frac{1}{2}\bar\omega^2\right)\delta\left(\bar\omega-\frac{2}{1+\gamma^2\theta^2}\right),\label{P_omega0}
\end{align}
where $\bar\omega\equiv\omega/\gamma\omega_0$, and $\omega_0$ is the fundamental cyclotron frequency in the rest frame with velocity $\beta\cos\psi$.
The radiation only occurs at the direction $\theta$ with 
\begin{align}
\gamma\theta=\left(\frac{2}{\bar\omega}-1\right)^{1/2}.
\end{align}
Notice that the particle's deflection angle $\psi$ only affects the normalized radiation power but not the typical frequency and the spectral shape. The typical radiation frequency is $\bar\omega\sim{\rm a~few}$ (corresponding to $\omega\sim\gamma\omega_0$).
According to Eq.(\ref{P_omega0}), for a certain viewing direction $\gamma\theta$, the emission is only at the frequency $\bar\omega=2/(1+\gamma^2\theta^2)$. Thus, \emph{the radiation spectrum of a single particle should be extremely narrow.} 
This result is consistent with the conclusion discussed in Section \ref{nspec}, that is, the narrow spectrum could be produced by that the particle’s acceleration is periodic during its radiation beam pointing to the observer.
If multiple radiating particles are distributed in a three-dimensional beam, the corresponding spectrum would become relatively wider than that of a single charged particle, but the spectrum is still narrow with $\Delta\nu/\nu\ll1$, see Appendix \ref{mechanism2a} for a detailed calculation. Therefore, if the observed narrow spectrum of FRBs is due to the intrinsic radiation mechanism by a single charged particle, it should be attributed to the scenario of $\gamma\psi\ll1$. 

The polarization properties for the scenario of $\gamma\psi\ll1$  are discussed in Appendix \ref{mechanism2a}. First, for the general polarization properties of the scenario with $\gamma\psi\ll1$, it can be easily proved that (1) If the acceleration is always on a straight line perpendicular to the particle's velocity, the polarization is fully linear; (2) If the acceleration rotates with a constant angular velocity on the plane perpendicular to the particle's velocity, the polarization is fully circular. In particle, for the small-pitch-angle synchrotron radiation by a single charged particle, high linear polarization (low circular polarization) only occur at $\bar\omega\sim1$ and $\gamma\theta\sim1$, otherwise, high circular polarization (low linear polarization) is dominant. For multiple radiating particles uniformly distributed in a three-dimensional beam, the observed linear and circular polarization degrees depend on the gyration directions of the radiating particles and whether the viewing direction is within the beaming angle. If all radiating particles have the same gyration directions, the polarization of this radiation mechanism would be almost 100\% circular polarization, otherwise, their polarizations would cancel out.

\subsection{Different astrophysical scenarios}\label{picture}

\begin{figure}
    \centering
    \includegraphics[width = 0.9\linewidth, trim = 90 50 70 50, clip]{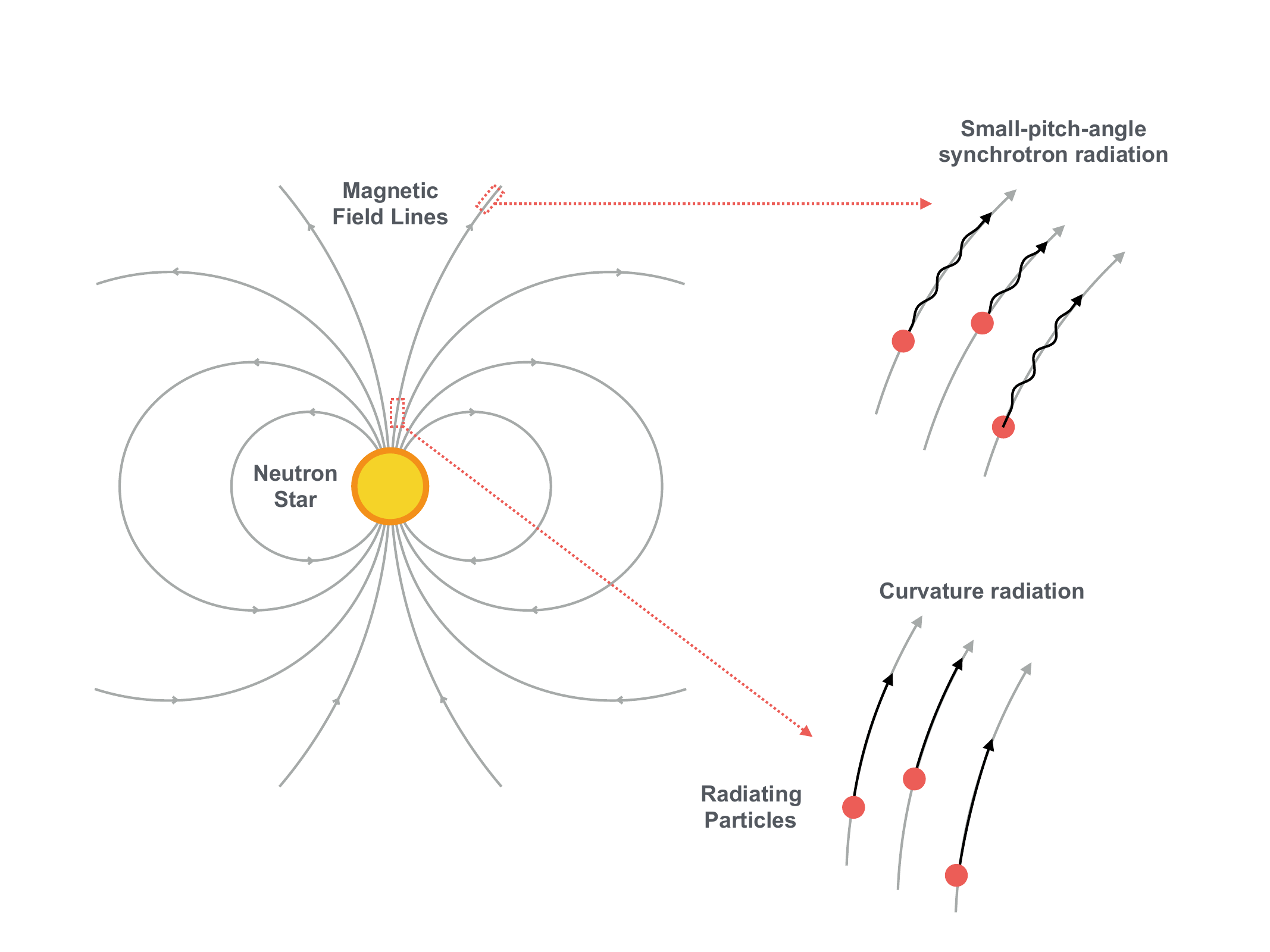}
    \caption{Two radiation mechanisms with $\gamma\psi\gg1$ and $\gamma\psi\ll1$ in the magnetosphere of a neutron star. In the inner region with a relatively strong magnetic field, the radiation mechanism is the curvature radiation. In the outer region with a relatively weak magnetic field, the radiation mechanism is the small-pitch-angle synchrotron radiation.}\label{magnetosphere} 
\end{figure}

The above two general radiation processes with $\gamma\psi\gg1$ and $\gamma\psi\ll1$ could appear in various astrophysical scenarios, e.g, in the magnetosphere of a neutron star or in a magnetized shocked medium. 
The radiation processes in the magnetosphere are shown in Figure \ref{magnetosphere}. In the inner region of the magnetosphere, due to the strong magnetic field, the electrons  move almost along the curvature field lines and produce curvature radiation\footnote{The accelerations of the electrons are essentially by the Lorentz forces of the drift velocities perpendicular to the field lines.}. In the picture of the curvature radiation, since the deflection angle (i.e., the deflection angle of the field line) is much larger than the radiation beaming angle, the condition of $\gamma\psi\gg1$ is satisfied.
In the outer region of the magnetosphere with the relatively weak magnetic field, the electrons would move in the field lines with a spiral trajectory and the corresponding radiation mechanism is the small-pitch-angle synchrotron radiation. The deflection angle $\psi$ in this case corresponds to the pitch angle of the synchrotron radiation, leading to $\gamma\psi\ll1$. One should notice the difference in the deflection angles $\psi$ in the above two scenarios due to the different mechanisms.

The critical conditions between the curvature radiation and the small-pitch-angle synchrotron radiation could be obtained as follows: when a charged particle with a Lorentz factor $\gamma$ moving along a trajectory with a curvature radius of $\rho$, the observer will see the radiation with the emission cone of angular width $1/\gamma$ around the observer direction and the typical timescale of the radiating process is $\tau_r=\rho/\gamma c$. 
The gyration period of an electron under a magnetic field is $\tau_B=2\pi/\omega_B=2\pi\gamma m_ec/eB$. If the radiation process is dominated by the small-pitch-angle synchrotron, the number of times for the electron's gyration during the time of $\tau_r$ must be much larger than once, which requires that the gyration period of the electron is much shorter than $\tau_r$, $\tau_B<\tau_r$, leading to the first necessary condition:
\begin{align}
B>B_{\rm cr,1}=\frac{2\pi\gamma^2m_ec^2}{e\rho}\simeq1.1~{\rm G}~\gamma_2^2\rho_8^{-1}.\label{cr1}
\end{align}
According to the Larmor formula, the radiation power of the small-pitch-angle synchrotron radiation is 
\begin{align}
P=\frac{2}{3}\frac{e^4\gamma^2B^2\beta_\perp^2}{m_e^2c^3}\simeq\frac{2}{3}\frac{e^4\gamma^2B^2\psi^2}{m_e^2c^3},
\end{align}
where $\beta_\perp=\beta\sin\psi\simeq\psi$ for $\beta\sim1$ and $\psi\ll1$. The cooling timescale is estimated by $P t_{\rm cool}\sim\gamma m_ec^2$, and one obtains
\begin{align}
t_{\rm cool}\sim\frac{3m_e^3c^5}{2e^4\gamma B^2\psi^2}\simeq5.2\times10^{10}~{\rm s}~\gamma_2^{-1}B_1^{-2}\psi_{-3}^{-2}.
\end{align}
If the small-pitch-angle synchrotron radiation is not significantly cooling during the electron moving along the field line, the cooling timescale $t_{\rm cool}$ must be larger than $\tau_r$, leading to the second necessary condition:
\begin{align}
B<B_{\rm cr,2}=\fraction{3m_e^3c^6}{2e^4\rho\psi^2}{1/2}\simeq3.9\times10^8~{\rm G}~\rho_8^{-1/2}\psi_{-3}^{-1}.\label{cr2}
\end{align}
Based on Eq.(\ref{cr1}) and Eq.(\ref{cr2}), the small-pitch-angle synchrotron radiation finally requires the conditions:
\begin{align}
1.1~{\rm G}~\gamma_2^2\rho_8^{-1}\lesssim B\lesssim3.9\times10^8~{\rm G}~\rho_8^{-1/2}\psi_{-3}^{-1}.\label{SPA_region}
\end{align}
Accordingly, the condition of the curvature radiation is
\begin{align}
B\gtrsim3.9\times10^8~{\rm G}~\rho_8^{-1/2}\psi_{-3}^{-1}.
\end{align}
As we expected, the small-pitch-angle synchrotron radiation is preferred to be in the outer region of the magnetosphere, and the curvature radiation is preferred to be in the inner region of the magnetosphere. 

Next, we make some comments about the observable properties of the small-pitch-angle synchrotron radiation in the magnetosphere. 
As discussed in Section \ref{mechanism2}, according to Eq.(\ref{freq10}) the typical radiation frequency of the small-pitch-angle synchrotron radiation is
\begin{align}
\nu\sim\gamma^2\nu_B=\frac{\gamma eB}{2\pi m_ec}\simeq2.8~{\rm GHz}~\gamma_2B_1.\label{spf}
\end{align}
Thus, the small-pitch-angle synchrotron radiation in the outer magnetosphere could be emitted at the GHz band
\footnote{Notice that the typical frequency at GHz band requires the magnetic field to be $B\sim10~{\rm G}$ according to Eq.(\ref{spf}). Such a weak field strength leads to the local magnetic energy density being much smaller than the plasma energy density, $B^2/8\pi\ll \xi^{-1}L_{\rm iso}/4\pi r^2c$, where $L_{\rm iso}$ is the isotropic luminosity of an FRB and $\xi$ is the radiation efficiency. This means that the radiating particles will straighten the field lines (that is similar to the magnetic field geometry of stellar wind), leading to a larger curvature radius $\rho$ than that of the dipole field at the same position. Such a process makes the small-pitch-angle radiation easy to produce because the field lines become almost parallel to the particles' velocities. However, how to generate coherent radiation is an issue for such a scenario, which need to be further analyzed in detail.}.
The polarization properties of the small-pitch-angle synchrotron radiation depend on the gyration motions of the charged particles in the magnetosphere, see the following discussions in Appendix \ref{mechanism2a}.
Before the charged particles enter the magnetosphere's outer region, in addition to the parallel velocities along the field line, the charged particles have a small drift velocity perpendicular to the field lines, providing an additional Lorentz force to make the particles move along the curved paths. Since the direction of the drift velocity only depends on the curved field line, the charged particles tend to have the same gyration direction when they enter the outer region of the magnetosphere. Thus, the small-pitch-angle radiation is expected to be highly circularly polarized (see Appendix \ref{mechanism2a}). Besides, since the spectral shape and the typical frequency of the small-pitch-angle radiation is independent of the particle's pitch angle, for multiple particles, the spectral bandwidth is mainly determined by the distribution of the particles' Lorentz factors.

\begin{figure}
    \centering
    \includegraphics[width = 0.9\linewidth, trim = 120 50 100 50, clip]{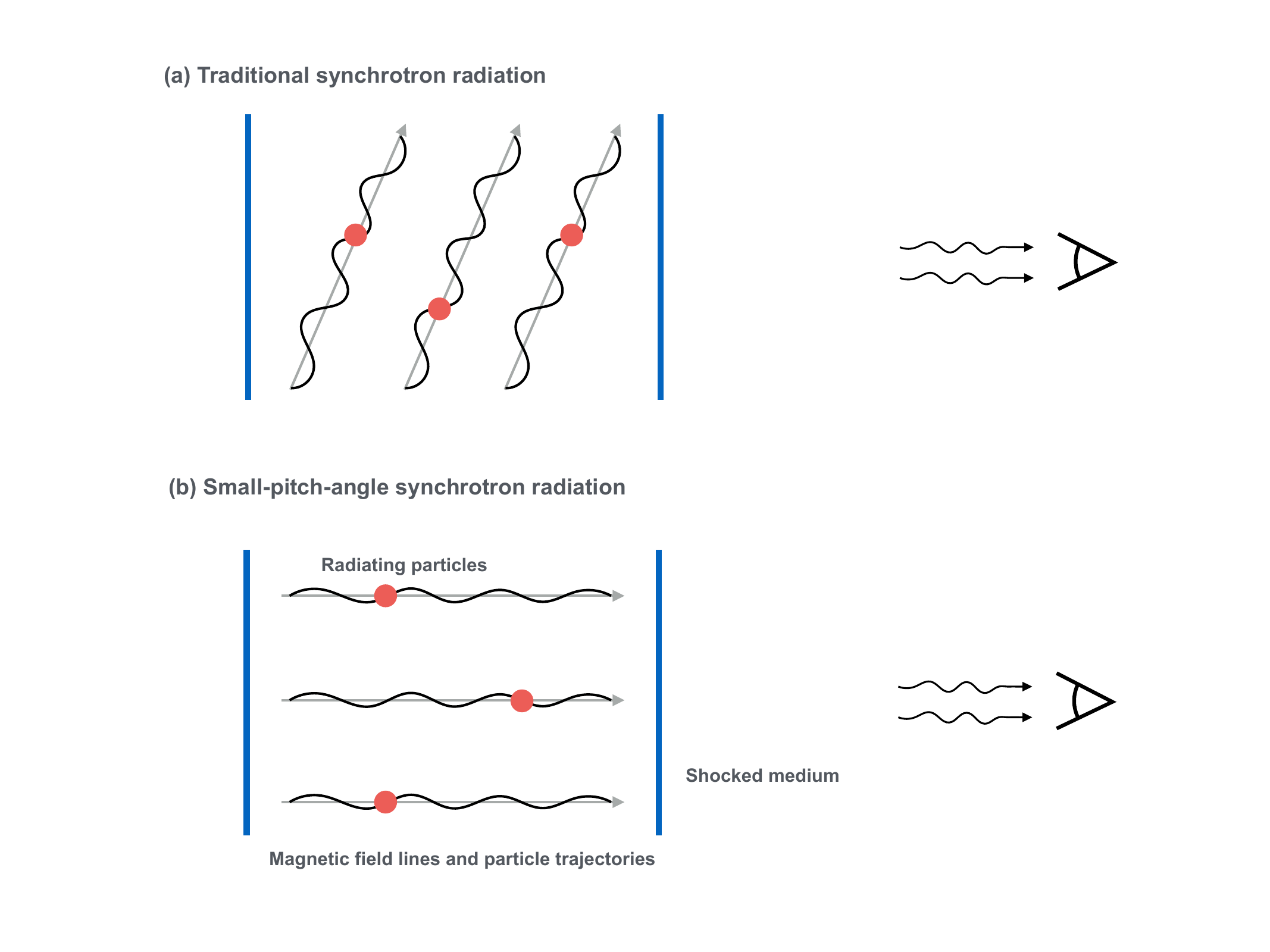}
    \caption{Two radiation mechanisms with $\gamma\psi\gg1$ and $\gamma\psi\ll1$ in the magnetized shocked medium. Panel (a) corresponds to traditional synchrotron radiation. Panel (b) corresponds to the small-pitch-angle synchrotron radiation, in which scenario, the emission region has a magnetic field almost parallel to the line of sight and the direction of the particles' injection is almost parallel to the field lines.}\label{shock}
\end{figure}

The above two general radiation processes with $\gamma\psi\gg1$ and $\gamma\psi\ll1$ might also occur in the magnetized shocked medium, which corresponds to the traditional synchrotron radiation and small-pitch-angle synchrotron radiation, as shown in the panel (a) and the panel (b) of Figure \ref{shock}. The critical condition between the two scenarios depends on the relative directions between the magnetic field, the particles' injection, and the viewing direction. If the direction of the particles' injection is almost parallel to the field lines, the radiation mechanism would be the small-pitch-angle synchrotron radiation, as shown in the panel (b) of Figure \ref{shock}. However, due to the shock compression, the magnetic field in the shocked medium usually has a significant component parallel to the shock surface. Thus, the small-pitch-angle injection process seems to be fine-tuning. Besides, in the magnetized shocked medium, since the directions of the particles' gyration motion are random, the small-pitch-angle synchrotron would be significantly depolarized for multiple particles.

\section{Radiation Spectra by Coherent Processes of Multiple Particles}\label{Section_Coherence}

In this section, we discuss how the coherent processes generate narrow spectra, including the bunching mechanism (Section \ref{Coherence_B}) and maser mechanism (Section \ref{Coherence_M}).

\subsection{Narrow spectra by bunching mechanism}\label{Coherence_B}

Coherent curvature radiation by charged bunches has been proposed as one of the popular ideas to explain the emission of FRBs \citep[e.g.,][]{Katz14,Katz18,Yang18,Yang23,Kumar20,Lu20,Cooper21}.
Compared with the spectrum by a single charged particle with $\gamma\psi\gg1$, see Section \ref{mechanism1} and Figure \ref{CR_spectrum}, a relatively narrow spectrum could be generated by the coherent curvature radiation from a structured bunch as proposed by \citet{Yang20} and \citet{Yang23} because the frequency structure of the burst is the Fourier transform of the spatial structure of the radiating charge density as pointed out by \citet{Katz18}, but it is still hard to explain the observed extremely narrow spectrum of some FRBs, e.g., FRB 20190711 with $\Delta\nu/\nu\sim0.05$. In particular, the dynamic fluctuation of the bunches would also make the spectrum show a white noise, especially at the low-frequency band \citep{Yang23}.
Since the narrow spectrum implies that the electromagnetic signal should be quasi-sinusoid in the short term as pointed out in Section \ref{nspec}, one possibility to generate a narrow spectrum is that the radiating bunches moving along the same trajectory are quasi-periodically distributed due to some special processes, like quasi-monochromatic Langmuir wave or oscillating pair creation in the charge-starved region. In this case, the radiation could be amplified at some harmonic frequencies, see the following discussion. 

We consider that the charged bunch distribution is quasi-periodic in the time domain, and the medium in the emission contains $N$ radiating bunches. Each radiating bunch emits a pulse of $E_0(t)$ with the same shape but with different arrival times. Thus, the total electric field from the multiple radiating bunches is given by
\begin{align}
E(t)=\sum_j^N E_0(t-t_j).
\end{align}
According to the time-shifting property of the Fourier transform, the total power spectrum of multiple radiating bunches is
\begin{align}
|E(\omega)|^2
=|E_0(\omega)|^2\left|\sum_j^Ne^{i\omega t_j}\right|^2,\label{coherence}
\end{align}
where $|E_0(\omega)|^2$ corresponds to the power spectrum of the first radiating bunch.
The coherence properties of the radiation by the multiple bunches are determined by the factor of $|\sum_j^N\exp(i\omega t_j)|^2$. 
If the multiple bunches are quasi-periodically distributed, one would have
\begin{align}
i\omega t_j=ij\omega/\omega_m+i\delta\phi_j,\label{cohphase}
\end{align}
where $1/\omega_m={\rm const.}$ corresponds to the period of the bunch distribution, and $\delta\phi_j$ corresponds to the relative random phase related to the phase of $j\omega/\omega_m$.

\begin{figure}
    \centering
    \includegraphics[width = 0.9\linewidth, trim = 0 0 0 0, clip]{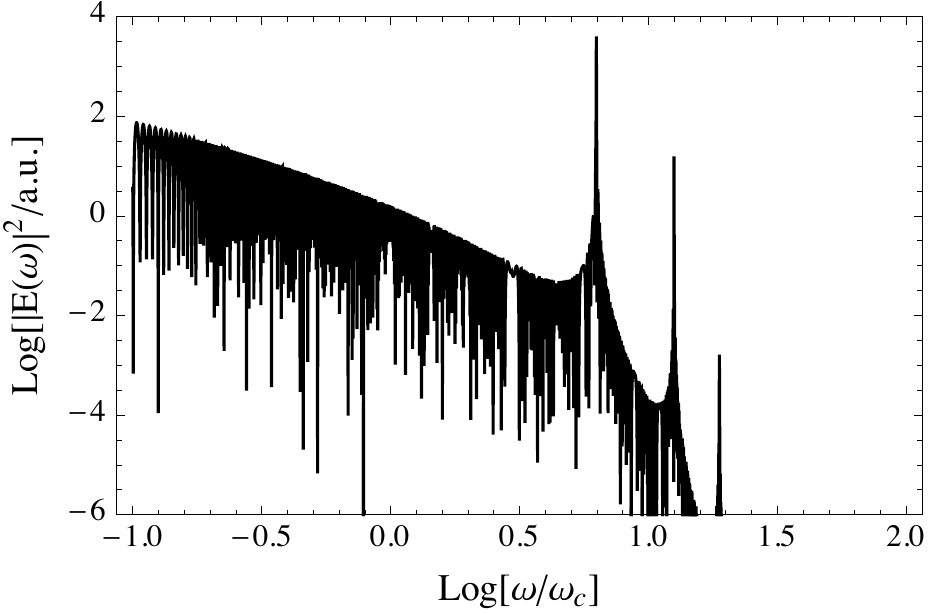}
    \includegraphics[width = 0.9\linewidth, trim = 0 0 0 0, clip]{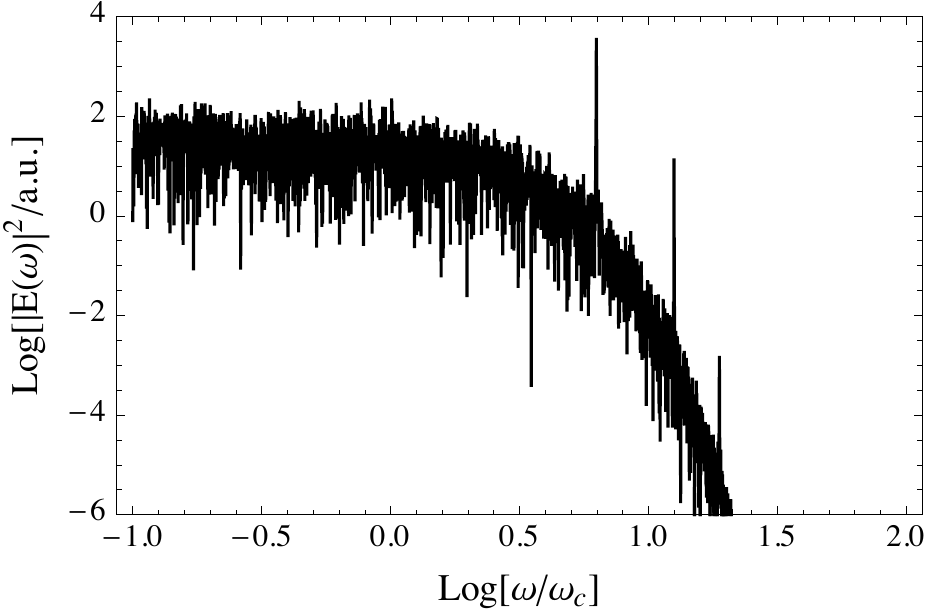}
    \caption{The spectra of multiple radiating bunches with a periodic or quasi-periodic distribution in the time domain. The top panel corresponds to a strictly periodic distribution with $\delta\phi_j=0$. The bottom panel corresponds to a quasi-periodic distribution with $\delta\phi_j$ uniformly distributed in $[-0.5,0.5]$.
    A curvature-radiation-like spectrum for a single point-source bunch, $|E_0(\omega)|^2\propto\omega^{2/3}\exp(-\omega/\omega_c)$, is taken as an example in this figure. We take $\omega_m=\omega_c$ and $N=10^3$. The peak frequency is at $2n\pi\omega_m$ with $n\in\mathbb{Z}^+$.}\label{periodic_particle} 
\end{figure}

We first consider the case of $\delta\phi_j=0$, which implies that the bunch distribution is strictly periodic.
Defining $z=\exp(i\omega/\omega_m)$, the modulus square of the sum of the phase factor in Eq.(\ref{coherence}) is calculated by
\begin{align}
\left|\sum_{j=1}^N z^j\right|^2&=\left|z\frac{z^N-1}{z-1}\right|^2=\frac{2-z^N-z^{*N}}{2-z- z^*}\nonumber\\
&=\frac{1-\cos(N\omega/\omega_m)}{1-\cos(\omega/\omega_m)}=\frac{\sin^2(N\omega/2\omega_m)}{\sin^2(\omega/2\omega_m)},
\end{align}
where $z^*$ is the conjugation of the complex number $z$, and the geometric sequence summation is used in the above calculation.
Therefore, the power spectrum by the multiple bunches with a periodic distribution is given by
\begin{align}
|E(\omega)|^2
=|E_0(\omega)|^2\sin^2\left(\frac{N\omega}{2\omega_m}\right)\sin^{-2}\left(\frac{\omega}{2\omega_m}\right).\label{PP_spectrum}
\end{align}
The radiation is coherently amplified when $\sin^2(\omega/2\omega_m)\sim0$, leading to the coherent peak frequencies at $\omega/\omega_m=2n\pi$ with $n\in\mathbb{Z}^+$. In the top panel of Figure \ref{periodic_particle}, we plot the power spectrum of the multiple radiating bunches with a periodic distribution according to Eq.(\ref{PP_spectrum}). The spectrum of a single bunch is assumed to be $|E_0(\omega)|^2\propto\omega^{2/3}\exp(-\omega/\omega_c)$ (corresponding to the spectrum of the curvature radiation by a single point-source bunch, see \citet{Yang18,Yang23}) as an example, and $N=10^3$ and $\omega_m=\omega_c$ are taken. We can see that the coherent radiation energy is radiated into multiples of $2\pi\omega_m$ with narrow bandwidths. 

Next, we consider that the bunch distribution is quasi-periodical with $\delta\phi_j\neq0$. We assume that the relative random phases $\delta\phi_j$ are uniformly distributed in a range of $[-\delta\phi_m,\delta\phi_m]$ with $0<\delta\phi_m\leqslant\pi$, and make a Monte Carlo simulation to calculate the total radiation based on Eq.(\ref{coherence}) and Eq.(\ref{cohphase}). In the bottom panel of Figure \ref{periodic_particle}, we plot the power spectrum of the multiple radiating bunches with a quasi-periodic distribution with $\delta\phi_m=0.5$. The other parameters are the same as that in the case of strictly periodic distribution. We can see that the coherent radiation energy is still radiated into multiples of $2\pi\omega_m$ with narrow bandwidths even for $\delta\phi_j\neq0$. Meanwhile, for a given frequency band, the continuous part of the spectrum becomes harder for a larger value of $\delta\phi_m$. In particular, if $\delta\phi_m=\pi$, the distribution of the multiple bunches would be random and the total power spectrum completely incoherent.

\subsection{Narrow spectra by maser mechanism}\label{Coherence_M}

In addition to the bunching mechanism, some maser mechanisms have been also proposed to be the radiation mechanism of FRBs \citep[e.g.,][]{Lyubarsky14,Metzger19,Waxman17,Beloborodov20}.
The maser mechanism corresponds to the negative absorption by radiating particles with inverted populations. 
A negative optical depth of $\tau_\nu$ at frequency $\nu$ causes an amplification by a factor of 
\begin{align}
A(\nu)=\exp(|\tau_\nu|). \label{maserintensity} 
\end{align}
Thus, the amplification involved in the maser mechanism could be very large even for an intermediate negative optical depth. 

We first consider that the electron distribution (i.e., the electron number density in an interval of $(\gamma,\gamma+d\gamma)$ with $\gamma$ as the electron Lorentz factor) is 
\begin{align}
n(\gamma)d\gamma\propto\gamma^p d\gamma\label{densitymaser},
\end{align} 
the radiation power of a single electron is $P(\nu,\gamma)$, and the scale of the emission region is $L$. The optical depth $\tau_\nu$ could be generally given by\footnote{Notice that this equation is very general because it is directly derived from Einstein coefficient. However, we should emphasize that this equation potentially assumes that the electron distribution is isotropic. If the electron distribution is one-dimensional (e.g., relativistic electrons and positrons moving along the magnetic field lines in the magnetosphere of a neutron star), the integrand in Eq.(\ref{negtau}) would be proportional to $dn(\gamma)/d\gamma$ \citep{Melrose78} rather than $\gamma^2\partial[n(\gamma)/\gamma^2]/\partial\gamma$, and the condition for the negative absorption would become $p>0$ rather than $p>2$.} \citep[e.g.,][]{Rybicki86}
\begin{align}
\tau_\nu=-\frac{L}{8\pi m_e\nu^2}\int P(\nu,\gamma)\left[\gamma^2\frac{\partial}{\partial\gamma}\left(\frac{n(\gamma)}{\gamma^2}\right)\right]d\gamma.\label{negtau}
\end{align}
Thus, the negative absorption requires that $\tau_\nu<0$, i.e., $\partial (n(\gamma)/\gamma^2)/\partial\gamma>0$, leading to the condition of 
\begin{align}
p>2.
\end{align}
In general, the electron distribution $n(\gamma)$ in the emission is usually wide and appears more complex than a single power-law distribution given by Eq.(\ref{densitymaser}), that is, the power-law index $p$ could be energy-dependent, $p(\gamma)$. In this case, the effective electrons contributing to the maser mechanism are only in the range in which $p(\gamma)>2$. We rewrite Eq.(\ref{negtau}) as
\begin{align}
\tau_\nu=\tau_{\nu,+}+\tau_{\nu-},
\end{align}
where $\tau_{\nu,+}$ and $\tau_{\nu,-}$ are the positive and negative optical depths with the integral ranges in Eq.(\ref{negtau}) corresponding to $p(\gamma)<2$ and $p(\gamma)>2$, respectively. For an effective maser mechanism, one usually has $|\tau_{\nu,-}|\gg\tau_{\nu,+}$ and $\tau_\nu\simeq\tau_{\nu,-}$ unless the electron distribution is very fine-tuning leading to a small net negative absorption. Thus, in the following discussion, we only consider the electrons distributed in 
\begin{align}
\gamma_{\min}<\gamma<\gamma_{\max}~~~\text{in which range}~p(\gamma)\sim p>2,
\end{align}
where $p(\gamma_{\min})=p(\gamma_{\max})\simeq2$.

\begin{figure}
    \centering
    \includegraphics[width = 0.9\linewidth, trim = 0 0 0 0, clip]{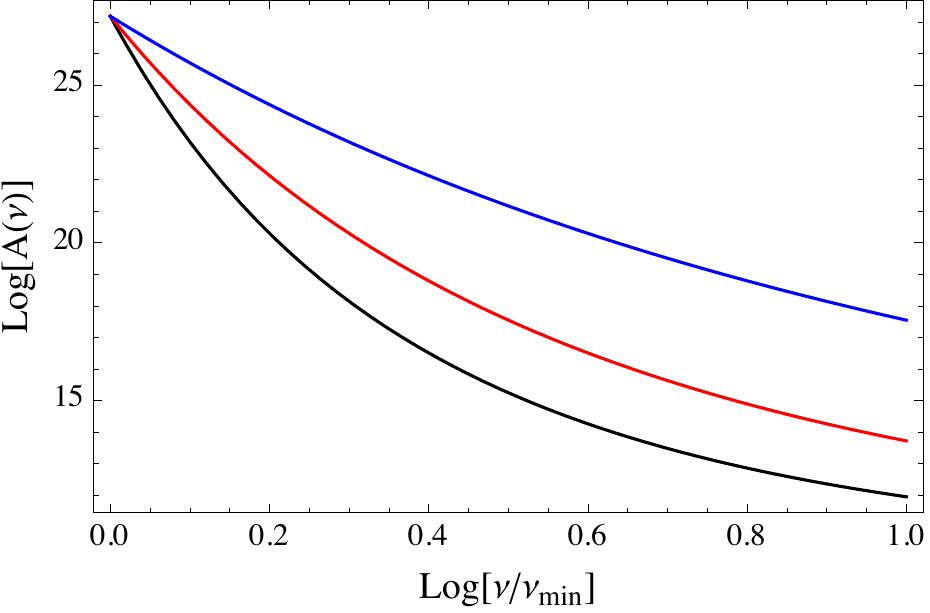}
    \caption{The amplification of the synchrotron maser mechanism as a function of frequency. The black, red and blue lines correspond to the electron distribution index $p=2.5,3,3.5$, respectively. The optical depth at $\nu_{\min}$ is assumed to be $\tau_{\nu,0}=10$ here. The frequency is dimensionless with $\nu_{\min}=\gamma_{\min}^2\nu_B$. Notice that the amplification becomes zero for $\nu<\nu_{\min}$ in this figure, see the text for the discussion.}\label{maser_amp} 
\end{figure}

We are mainly interested in the synchrotron maser that has been proposed as one of the popular ideas to explain the coherent emission of FRBs \citep[e.g.,][]{Lyubarsky14,Metzger19,Waxman17,Beloborodov20}. The $\delta$-function approximation for the synchrotron emissivity can be written as
\begin{align}
P(\nu,\gamma)&\simeq\frac{4}{3}c\sigma_{\rm T}\frac{B^2}{8\pi}\gamma^2\delta(\nu-\gamma^2\nu_B)\nonumber\\
&=\frac{4\pi}{9}\frac{e^3B}{m_ec^2}\gamma\delta\left(\gamma-\sqrt{\frac{\nu}{\nu_B}}\right),\label{Psyn}
\end{align}
where $\nu_B=eB/2\pi m_ec$ is the cyclotron frequency. Substituting Eq.(\ref{densitymaser}) and Eq.(\ref{Psyn}) in Eq.(\ref{negtau}) gives
\begin{align}
\tau_\nu&\sim\frac{e^3BL}{18m_e^2c^2}\frac{n(\gamma\sim\sqrt{\nu/\nu_B})}{\nu^2}\nonumber\\
&\propto\nu^{(p-4)/2}.~~~\text{for}~\gamma_{\min}^2\nu_B<\nu<\gamma_{\max}^2\nu_B.\label{masertau}
\end{align}
Based on Eq.(\ref{maserintensity}) and Eq.(\ref{masertau}), we take the form of amplification as $A(\nu)=\exp[\tau_{\nu_0}(\nu/\nu_{\min})^{(p-4)/2}]$, where $\tau_{\nu,0}$ corresponds to the optical depth at $\nu_{\min}=\gamma_{\min}^2\nu_B$ and $2<p<4$ is assumed. In Figure \ref{maser_amp}, we plot the spectra of the synchrotron maser, and an intermediate optical depth of $\tau_{\nu,0}=10$ is assumed. The black, red, and blue lines correspond to $p=2.5,3,3.5$, respectively. We can see that for these typical parameters, the amplification varies several orders of magnitude within a narrow bandwidth. Thus, the spectrum of synchrotron maser should be extremely narrow.

\section{Radiation spectra corrected by radiative transfer processes}\label{transfer}

The radiative transfer processes include absorption (Section \ref{absorption}) and scattering (Section \ref{scattering}). Since both absorption and scattering are dominant in some dense environments, they are important only at the circumburst medium or the emission region.

\subsection{Absorption}\label{absorption}

The absorption coefficients $\alpha_\nu$ of most absorption processes (e.g., free-free absorption, synchrotron absorption, plasma absorption, etc.) have a negative correlation with the frequency of electromagnetic waves, $\alpha_\nu\propto\nu^{-\delta}$ with $\delta>0$, that is, the lower the frequency, the more significant the absorption effect. We consider that the incident intensity is $I_{\nu,0}$ and the scale of the absorption region is $L$, then the outgoing intensity has a spectrum of $I_{\nu,0}\exp(-\alpha_\nu L)$. If the frequency $\nu$ is much less than the absorption frequency $\nu_{\rm abs}$ defined by $\tau_\nu(\nu_{\rm abs})=\alpha_\nu(\nu_{\rm abs})L=1$, the intensity would be significantly cut off, i.e., $I_{\nu}(\nu<\nu_{\rm abs})\ll I_{\nu,0}$. This result suggests that the radio bursts emitted during a limited time of $T$ would have the same low-frequency cutoff at $\nu_{\rm low}\sim\nu_{\rm abs}$ if the intrinsic spectrum is intrinsically wide and the typical evolution timescale of the absorption parameters is much longer than $T$. However, such a conclusion is inconsistent with the observations of some FRB repeaters, suggesting that the absorption effects do not significantly change the observed spectra. 

For example, FRB 200428 from Galactic magnetar SGR J1935+2154 consisted of two sub-bursts separated by $T\sim 29~{\rm ms}$ \citep{CHIME20}. The first component did not show a significant low-frequency cutoff at the telescope's bandwidth, implying that the low-frequency cutoff satisfies $\nu_{\rm low}\ll400~{\rm MHz}$, and the second component had a low-frequency cutoff at $\nu_{\rm low}\sim(400-500)~{\rm MHz}$. 
The relative variation of the low-frequency cutoffs, $\delta\nu_{\rm low}/\nu_{\rm low}$, was several tens percent during $T\sim 29~{\rm ms}$. This observation suggests that the relative variations of the environment parameters (e.g., average electron density, absorption region lengthscale, temperature, magnetic field, etc.) in the absorption region also reached several tens percent during such an extremely short time, which is obviously not realistic.
Therefore, the low-frequency cutoffs of the observed spectra of some FRB sources should not be attributed to the absorption processes.

\subsection{Electron Scattering}\label{scattering}

The most important mechanism of various scatterings is electron scattering, including relativistic and nonrelativistic types. Relativistic electron scattering mainly refers to inverse Compton scattering. For a general large-angle scattering process by a relativistic electron with Lorentz factor of $\gamma$, the inverse Compton scattering process coverts a photon with frequency $\nu$ to a high-frequency one with frequency $\sim\gamma^2\nu$. \emph{Since the cross section (i.e., Thomson cross section) is independent of the photon frequency, the scattering process simultaneously suppresses the specific intensity in all bands, leading to the spectral shape of the unscattered radiation unchanged}. 

The nonrelativistic electron scattering is called Thomson scattering. As an elastic scattering process, the total amount of radiation emitted per unit frequency is almost equal to the total amount absorbed in that same frequency and the cross section is also independent of frequency. 
However, repeating scattering process can build up a substantial effect, i.e., induced Compton scattering\footnote{Induced Compton scattering could be significant for relativistic electron scattering when the scattering angle is so small that the scattering frequency is close to the incident one.}. The induced Compton scattering on relatively dense plasma would unavoidably affect the emergent radiation. We consider that the electron number density is $n_e$, the electron temperature is $T$, the photon occupation number (i.e, the average number of photons in a state) is \citep[e.g.,][]{Rybicki86,Wilson82,Yang23}
$n_\gamma=kT_B/h\nu=(c^2/2h\nu^3)I_\nu$,
where $T_B$ is the radiation brightness temperature and $I_\nu$ is the radiation intensity. The kinetic equation for the photon interacting with nonrelativistic electrons is \citep{Kompaneets57,Syunyaev71,Rybicki86}
\begin{align}
\frac{\partial n_\gamma}{\partial t}=\frac{h\sigma_{\rm T} n_e}{m_ec}\frac{1}{\nu^2}\frac{\partial}{\partial\nu}\nu^4\left(n_\gamma^2+n_\gamma+\frac{kT}{h}\frac{\partial n_\gamma}{\partial \nu}\right).
\end{align}
This is the so-called Kompaneets equation, which describes the evolution of the photon distribution function due to repeated nonrelativistic electron scattering. 
For FRBs with extremely temperatures, we are only interested in $n_\gamma\gg1$ (i.e., $kT_B\gg h\nu$) and $n_\gamma\gg kT/h\nu$ (i.e., $T_B\gg T$), leading to the second and third terms in the parentheses ignored.
Thus, the kinetic equation could be written as
\begin{align}
\frac{\partial n_\gamma}{\partial t}=\frac{2h\sigma_{\rm T}n_en_\gamma}{m_ec}\frac{\partial(\nu^2 n_\gamma)}{\partial\nu}.
\end{align}

According to the above equation, the radiation with the brightness temperature $T_B$ is unable to escape from a region with an effective optical depth $\tau_{\rm ind}=(kT_B/m_ec^2)\tau_{\rm T}>1$ \citep{Lyubarsky16}, where $\tau_{\rm T}$ is the optical depth of Thomson scattering.
The total number of photons is conserved and the photons are redistributed toward lower frequencies, which leads a larger brightness temperature $T_B$ so that the rate of redistribution increases further out. Finally, the photons are eventually absorbed by some absorption processes, e.g., free-free absorption, synchrotron absorption, plasma absorption, etc.
In conclusion, the induced Compton scattering makes the emergent spectrum softer than the initial incident spectrum and cuts off the spectrum at a low frequency determined by the absorption processes. Based on the discussion in Section \ref{absorption}, since the radio bursts of an FRB repeater did not show the same low-frequency cutoff during a limited time, the effect of induced Compton scattering should also be ignored.

The above electron scattering mechanisms potentially assume that the scattering processes are linear, which means that the electron oscillation is nonrelativisitc and the interaction between electromagnetic wave and electron is dominated by electric force (i.e., Lorentz force could be ignored due to the nonrelativistic motion of electrons). We consider an electromagnetic wave has an electric field $E$ and frequency $\omega$, and define the strength parameter as 
\begin{align}
a=\frac{v_{\rm os}}{c}=\frac{eE}{2\pi m_ec\nu},
\end{align}
where $v_{\rm os}=eE/2\pi m_e\nu$ is the typical oscillation velocity due to the electric force. For $a\ll1$, the electron oscillation is nonrelativistic and the interaction is dominated by the electric force, which means that the Thomson theory is valid. For $a\gg1$, the electron oscillation becomes relativistic and the interaction between electron and electromagnetic wave enters the nonlinear regime of ``strong waves''. In particular, the nonlinear effect makes the cross section enhanced by a factor of $a^2$ \citep[e.g.,][]{Sarachik70,Yang20b,Beloborodov22,Qu22}
\begin{align}
\sigma\sim a^2\sigma_{\rm T}\propto\frac{I_\nu}{\nu},\label{nonlinear_section}
\end{align}
where $I_\nu$ is the incident intensity.
Thus, the lower the frequency or the high the intensity, the more significant the scattering. According to Eq.(\ref{nonlinear_section}), the radio bursts would have low-frequency cutoff at $\nu_{\rm low}\propto I_\nu$, which is worthy to test the nonlinear effect by the observations of FRB repeaters. In addition to the escape unscattered photons, high-frequency photons would be produced by the relativistically oscillating electrons but they are significantly outside the band of the incident photons, which is beyond the scope of this work.
 
\section{Radiation spectra corrected by interference processes}\label{propagation}

The interference processes that can change the radiation spectra mainly include scintillation, gravitational lensing, and plasma lensing. Due to the wave interference, the spectra are coherently enhanced at some frequencies but coherently reduced at some others frequencies.

\subsection{Scintillation}

Generally, ``scintillation'' refers to the spectral modulations caused by the coherent combination of multiple waves, and ``temporal scattering'' refers to the temporal broadening of pulses due to the multipath propagation.
For a certain plasma screen, the relation between scintillation and temporal scattering is $\Delta\nu_{\rm sci}\tau_{\rm s}\sim1/2\pi$, where $\Delta\nu_{\rm sci}$ is the scintillation bandwidth and $\tau_{\rm s}$ is the scattering time. If the observed temporal scattering is $\tau_{\rm s}\sim1~{\rm ms}$, the corresponding scintillation bandwidth should be $\Delta\nu_{\rm sci}\sim160~{\rm Hz}$, which is much smaller than the observed scintillation bandwidth of $\Delta\nu_{\rm sci}\sim1~{\rm MHz}$. 
Therefore, for extragalactic FRBs, the observed scintillation is proposed to be mainly contributed by the interstellar medium within the Milky Way, and the observed scattering time is more likely contributed by the circumburst medium or the interstellar medium in the FRB host galaxy.

In this section, we consider that the spectral shape with a typical bandwidth $\gtrsim100~{\rm MHz}$ might be attributed to another plasma screen that is different from the regions contributing the observed temporal scattering and the narrow-bandwidth scintillation with $\Delta\nu_{\rm sci}\sim1~{\rm MHz}$.
The relatively wider scintillation bandwidth of $\Delta\nu_{\rm sci}\gtrsim100~{\rm MHz}$ requires that the screen should be close to the FRB source.
We consider that the plasma screen has a thickness of $\Delta R$, and the power spectrum of electron density fluctuations in the plasma screen is power-law with
\begin{align}
P(k)=C_N^2k^{-\beta},~~~{\rm for}~2\pi L^{-1}\lesssim k\lesssim 2\pi l_0^{-1},\label{power}
\end{align}
where $\beta$ is the spectral index of the three-dimensional power spectrum (Kolmogorov turbulence has $\beta=11/3$), $k=2\pi/l$ is the spatial wavenumber, $l_0$ and $L$ are the inner and outer scales of the inertial range of the turbulence, respectively, and the normalization factor $C_N^2$ is given by \citep[e.g.,][]{Xu17,Yang22}
\begin{align}
C_N^2\simeq
\left\{
\begin{aligned}
&\frac{3-\beta}{2(2\pi)^{4-\beta}}l_0^{3-\beta}\delta n_{e}^2,~&{\rm for}~\beta<3,\\
&\frac{\beta-3}{2(2\pi)^{4-\beta}}L^{3-\beta}\delta n_{e}^2,~&{\rm for}~\beta>3, 
\end{aligned}
\right.
\end{align}
where $\delta n_{e}^2$ is the total mean-squared density fluctuation.
We define the diffractive lengthscale $l_{\rm diff}$ that represents the transverse separation for which the root-mean-squared difference of the wave phases is equal to unit rad. For the electromagnetic wave with a wavelength of $\lambda$, the diffractive lengthscale $l_{\rm diff}$ is \citep[e.g.,][]{Xu17,Yang22}
\begin{align}
l_{\rm diff}=
\left\{
\begin{aligned}
&(f_{1,\alpha}\pi^2 r_e^2\lambda^2l_0^{\beta-4}C_N^2\Delta R)^{-\frac{1}{2}},~&{\rm for}~l_{\rm diff}<l_0,\\
&(f_{2,\alpha}\pi^2 r_e^2\lambda^2 C_N^2\Delta R)^{\frac{1}{2-\beta}},~&{\rm for}~l_{\rm diff}>l_0, 
\end{aligned}
\right.\label{ldiff}
\end{align}
where $r_e$ is the classical electron radius, $f_{1,\alpha}=\Gamma(1-\alpha/2)$, $f_{2,\alpha}=[\Gamma(1-\alpha/2)/\Gamma(1+\alpha/2)](8/\alpha 2^{\alpha})$, and $\alpha=\beta-2$. 
For the Kolmogorov turbulence with $\alpha=5/3~(\beta=11/3)$, one has $f_{1,\alpha}=5.6$ and $f_{2,\alpha}=8.9$, respectively.

The scattering angle of the electromagnetic waves is $\theta_{\rm s}\simeq\lambda/2\pi l_{\rm diff}$, and the transverse scale of the visible part of the plasma screen is \citep[e.g.,][]{Yang22}
\begin{align}
l_{\rm s}(\lambda)=\theta_{\rm s} R\simeq\frac{\lambda R}{2\pi l_{\rm diff}},\label{lsca}
\end{align}
where $R$ is the distance from the plasma screen to the FRB source.
The temporal scattering time could be then estimated by
\begin{align}
\tau_{\rm s}(\lambda)\simeq\frac{l_{\rm s}^2}{2Rc}=\frac{\lambda^2R}{8\pi^2 cl_{\rm diff}^2}.\label{tau}
\end{align}
Using Eq.(\ref{ldiff}) and $\Delta R\sim R$, the temporal scattering time satisfies
\begin{align}
\tau_{\rm s}(\lambda)\simeq
\left\{
\begin{aligned}
&\frac{3-\beta}{16(2\pi)^{4-\beta}}\frac{f_{1,\alpha}r_e^2}{cl_0}\delta n_{e}^2R^2\lambda^4,&l_{\rm diff}<l_0,\\
&\frac{(3-\beta)^{\frac{2}{\beta-2}}}{2^{\frac{\beta+4}{\beta-2}}}\frac{f_{2,\alpha}^{\frac{2}{\beta-2}}r_e^{\frac{4}{\beta-2}}}{cl_0^{\frac{2(\beta-3)}{\beta-2}}}\delta n_{e}^{\frac{4}{\beta-2}}R^{\frac{\beta}{\beta-2}}\lambda^{\frac{2\beta}{\beta-2}},&l_{\rm diff}>l_0. 
\end{aligned}
\right.\label{tau1}
\end{align}
for $\beta<3$ and 
\begin{align}
\tau_{\rm s}(\lambda)\simeq
\left\{
\begin{aligned}
&\frac{\beta-3}{16(2\pi)^{4-\beta}}\frac{f_{1,\alpha}r_e^2l_0^{\beta-4}}{cL^{\beta-3}}\delta n_{e}^2R^2\lambda^4,&l_{\rm diff}<l_0,\\
&\frac{(\beta-3)^{\frac{2}{\beta-2}}}{2^{\frac{\beta+4}{\beta-2}}}\frac{f_{2,\alpha}^{\frac{2}{\beta-2}}r_e^{\frac{4}{\beta-2}}}{cL^{\frac{2(\beta-3)}{\beta-2}}}\delta n_{e}^{\frac{4}{\beta-2}}R^{\frac{\beta}{\beta-2}}\lambda^{\frac{2\beta}{\beta-2}},&l_{\rm diff}>l_0. 
\end{aligned}
\right.\label{tau2}
\end{align}
for $\beta>3$.
The corresponding scintillation bandwidth is
\begin{align}
&\Delta\nu_{\rm sci}(\nu)\simeq\frac{1}{2\pi\tau_{\rm s}}\nonumber\\
&\simeq
\left\{
\begin{aligned}
&\frac{16(2\pi)^{3-\beta}}{3-\beta}\frac{l_0}{c^3f_{1,\alpha}r_e^2}\delta n_{e}^{-2}R^{-2}\nu^{4},&l_{\rm diff}<l_0,\\
&\frac{2^{\frac{\beta+4}{\beta-2}}}{2\pi (3-\beta)^{\frac{2}{\beta-2}}}\frac{l_0^{\frac{2(\beta-3)}{\beta-2}}}{c^{\frac{\beta+2}{\beta-2}}f_{2,\alpha}^{\frac{2}{\beta-2}}r_e^{\frac{4}{\beta-2}}}\delta n_{e}^{\frac{4}{2-\beta}}R^{\frac{\beta}{2-\beta}}\nu^{\frac{2\beta}{\beta-2}},&l_{\rm diff}>l_0. 
\end{aligned}
\right.\label{scint1}
\end{align}
for $\beta<3$ and 
\begin{align}
&\Delta\nu_{\rm sci}(\nu)\simeq\frac{1}{2\pi\tau_{\rm s}}\nonumber\\
&\simeq
\left\{
\begin{aligned}
&\frac{16(2\pi)^{3-\beta}}{\beta-3}\frac{L^{\beta-3}}{c^3f_{1,\alpha}r_e^2l_0^{\beta-4}}\delta n_{e}^{-2}R^{-2}\nu^{4},&l_{\rm diff}<l_0,\\
&\frac{2^{\frac{\beta+4}{\beta-2}}}{2\pi (\beta-3)^{\frac{2}{\beta-2}}}\frac{L^{\frac{2(\beta-3)}{\beta-2}}}{c^{\frac{\beta+2}{\beta-2}}f_{2,\alpha}^{\frac{2}{\beta-2}}r_e^{\frac{4}{\beta-2}}}\delta n_{e}^{\frac{4}{2-\beta}}R^{\frac{\beta}{2-\beta}}\nu^{\frac{2\beta}{\beta-2}},&l_{\rm diff}>l_0. 
\end{aligned}
\right.\label{scint2}
\end{align}
for $\beta>3$.
For the Kolmogorov turbulence with $\beta=11/3$, the typical value of the scintillation bandwidth is given by Eq.(\ref{scint2})
\begin{align}
\Delta\nu_{\rm sci}\simeq130~{\rm MHz}~L_{15}^{2/3}l_{0,13}^{1/3}\delta n_{e,3}^{-2}R_{15}^{-2}\nu_9^4,
\end{align}
where $l_{\rm diff}<l_0$ is satisfied for the typical parameters. Therefore, if the spectral shape with a typical bandwidth of $\Delta\nu_{\rm sci}\sim100~{\rm MHz}$ is attributed to the plasma screen, the environment near the FRB source would be required to be intermediately dense (e.g. $n_e\sim\delta n_e\sim10^{3}~{\rm cm^{-3}}$ at $R\sim10^{15}~{\rm cm}$) and turbulent. 

\subsection{Gravitational lensing}\label{gravitational_lens}

The time-delay effect of gravitational lensing can modulate the spectrum of a transient \citep{Gould92,Barnacka12}, which mainly depends on the mass and position of the lensing object. We consider the lensing object has a mass of $M$. The distances from the lens to the FRB source, from the observer to the lens, and from the observer to the FRB source are $d_{\rm ls}$, $d_{\rm ol}$ and $d_{\rm os}$, respectively. The Einstein radius of gravitational lensing is 
\begin{align}
r_{\rm E}=\left(\frac{4GM}{c^2}\frac{d_{\rm ol}d_{\rm ls}}{d_{\rm os}}\right)^{1/2}.
\end{align}
We define the source position projected onto the lens plane as $r_{\rm s}$, then the image positions are given by
\begin{align}
r_{\pm}=\frac{1}{2}\left(r_{\rm s}\pm\sqrt{r_{\rm s}^2+4r_{\rm E}^2}\right).
\end{align}
The time delay between the two images is
\begin{align}
\delta t=\frac{2GM}{c^3}\left[\left(\frac{r_+}{r_{\rm E}}\right)^2-\left(\frac{r_-}{r_{\rm E}}\right)^2+2\ln\left|\frac{r_+}{r_-}\right|\right].
\end{align}
The phase difference between the rays from the two images is
\begin{align}
\delta\phi=2\pi\nu\delta t.
\end{align}
A gravitational lensing event with remarkable spectral modulation requires $\delta\phi\sim1$ and $r_{\rm s}\lesssim r_{\rm E}$, leading to
\begin{align}
\delta t&=\frac{1}{2\pi\nu}\simeq1.6\times10^{-10}~{\rm s}~\nu_9^{-1},\\
M&\sim\frac{c^3\delta t}{G}\simeq2.0\times10^{-5}M_\odot\delta t_{-10}.
\end{align}
The time delay between two images is much shorter than the durations of FRBs, and the lens is required to be the planet-like object.
The amplitude contributed by the image at position $r_\pm$ is $\mathcal{A}_\pm\propto\exp(i\phi)/\sqrt{|1-r_{\rm E}^4/r_\pm^4|}$ \citep[e.g.,][]{Barnacka12}. The amplification is obtained by summing the amplitudes of the two images, $A\equiv|\mathcal{A}|^2=|\mathcal{A}_++\mathcal{A}_-|^2$, which is given by
\begin{align}
A&=\frac{1}{|1-r_{\rm E}^4/r_+^4|}+\frac{1}{|1-r_{\rm E}^4/r_-^4|}\nonumber\\
&+\frac{2\cos\delta\phi}{\sqrt{|1-r_{\rm E}^4/r_+^4|}\sqrt{|1-r_{\rm E}^4/r_-^4|}}.
\end{align}

\begin{figure}
    \centering
    \includegraphics[width = 0.9\linewidth, trim = 0 0 0 0, clip]{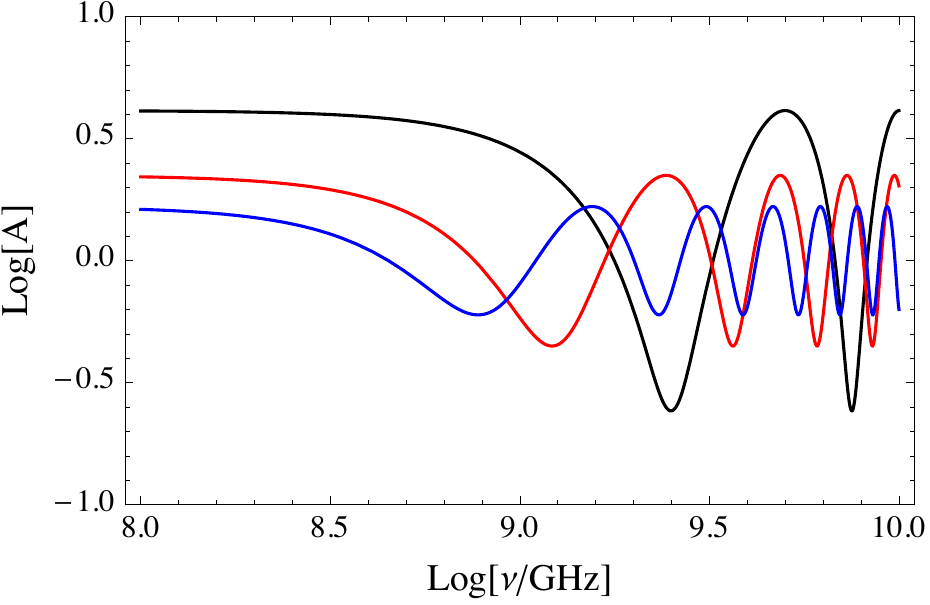}
    \includegraphics[width = 0.9\linewidth, trim = 0 0 0 0, clip]{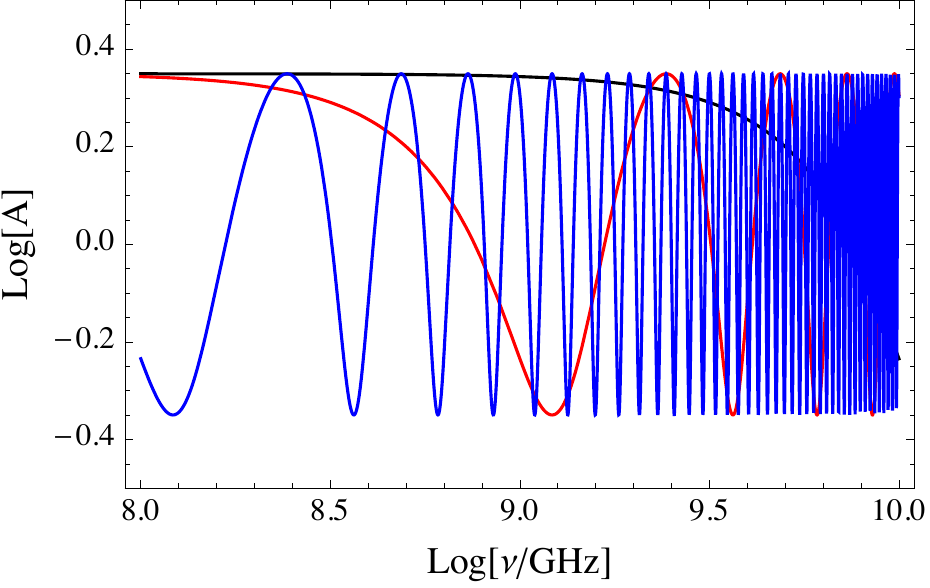}
    \caption{The amplification of the gravitational lensing as a function of frequency. The top panel takes the lens mass as $M=10^{-5}M_\odot$. The black, red, and blue lines correspond to $r_{\rm s}/r_{\rm E}=0.5,1.0,1.5$, respectively. The bottom panel takes the source projected position as $r_{\rm s}/r_{\rm E}=1.0$. The black, red, and blue lines correspond to the lens mass of $M=10^{-6}M_\odot,10^{-5}M_\odot,10^{-4}M_\odot$, respectively. The distances from the lens to the FRB source, from the observer to the lens, and from the observer to the FRB source are taken as $d_{\rm ls}=1~{\rm kpc}$, $d_{\rm ol}=1~{\rm Gpc}$ and $d_{\rm os}=1~{\rm Gpc}$, respectively.}\label{glens}
\end{figure}

In Figure \ref{glens}, we plot the amplification $A$ of the gravitational lensing as a function of frequency.  We take the distances as $d_{\rm ls}=1~{\rm kpc}$, $d_{\rm ol}=1~{\rm Gpc}$ and $d_{\rm os}=1~{\rm Gpc}$, respectively. The top panel takes the lens mass as $M=10^{-5}M_\odot$, and the black, red, and blue lines correspond to $r_{\rm s}/r_{\rm E}=0.5,1,1.5$, respectively. The bottom panel takes the source projected position as $r_{\rm s}/r_{\rm E}=1.0$. The black, red, and blue lines correspond to the lens mass of $M=10^{-6}M_\odot,10^{-5}M_\odot,10^{-4}M_\odot$, respectively.
We can see that spectral modulation becomes significant at GHz band only when $M\sim10^{-5}M_\odot$ and $r_{\rm s}\sim r_{\rm E}$ as pointed out above.

Although a planet-like object can generate the spectral modulation at the GHz band, the observed variation of the spectra of an FRB repeater seems not to support such a scenario due to the following reasons. 
For the above distance parameters, the typical lensing timescale is 
\begin{align}
t_{\rm lens}=\frac{r_{\rm E}}{v}\sim10^4~{\rm s}~M_{\odot,-5}v_7^{-1}d_{\rm os,Gpc}^{-1/2}d_{\rm ol,Gpc}^{1/2}d_{\rm ls,kpc}^{1/2}, 
\end{align}
where $v$ is the transverse velocity of the lens. This timescale is much shorter than the several-year active period of an FRB repeater. 
The observation of FRB repeaters showed that the burst-to-burst spectral variation is always present during the active periods. If the variation is due to the gravitational lensing, one would not see the persistent variation of the spectra during the long term, because the gravitational lensing event is most likely one-time. On the other hand, this timescale is much longer than the waiting time of an FRB repeater with a significant variation of spectra. It cannot explain the remarkable burst-to-burst variation of the spectra in an extremely short time, e.g., the spectrum variation of two components in FRB 200428 \citep{CHIME20}.
Therefore, the spectral modulation by gravitational lensing could be ruled out.

\subsection{Plasma lensing}

Plasma lensing plays an important role in the ``extreme scattering events'' as shown in the light curves of some radio pulsars and active galactic nuclei \citep{Fiedler87,Bannister16}. Similar to gravitational lensing discussed in Section \ref{gravitational_lens}, plasma lensing can also cause the time-delay effect. However, the time delay in plasma lensing is frequency-dependent due to the plasma dispersion \citep[e.g.,][]{Er20}, which would lead to a different spectral modulation compared to that of gravitational lensing. 

Following \citet{Cordes17}, we consider the plasma structure has a form with the dispersion measure satisfying ${\rm DM}(x)={\rm DM}_l\exp(-x^2/a^2)$, which yields a phase perturbation $\phi_\lambda=-\lambda r_e{\rm DM}(x)$, where $\lambda$ is the wavelength and $r_e$ is the classical electron radius. The distances from the lens to the FRB source, from the observer to the lens, and from the observer to the FRB source are $d_{\rm ls}$, $d_{\rm ol}$ and $d_{\rm os}$, respectively. The transverse coordinates in the source, lens, and observer's planes are $x_{\rm s}$, $x$ and $x_{\rm o}$, and one defines the dimensionless coordinates $u_{\rm s}=x_{\rm s}/a$, $u=x/a$ and $u_{\rm o}=x_{\rm o}/a$, respectively. The effective transverse offset can be expressed as $u'=(d_{\rm ol}/d_{\rm os})u_{\rm s}+(d_{\rm ls}/d_{\rm os})u_{\rm o}$. The lens equation in geometric optics gives \citep{Clegg98,Cordes17}
\begin{align}
u[1+\alpha\exp(-u^2)]=u',\label{cor1}
\end{align}
where the parameter $\alpha$ is given by
\begin{align}
\alpha=\frac{\lambda^2r_e{\rm DM}_l}{\pi a^2}\left(\frac{d_{\rm ol}d_{\rm ls}}{d_{\rm os}}\right).
\end{align}
There are either one or three solutions for $u$ for a given offset $u'$ based on Eq.(\ref{cor1}) \citep[see the detailed discussion in ][]{Cordes17}. Due to the limited spatial resolution of radio telescopes, the separated images might be extremely difficult to resolve. In the geometrical optics regime, the focusing or defocusing of incident wavefronts yields an amplification \citep{Clegg98,Cordes17}
\begin{align}
A=|1+\alpha(1-2u^2)\exp(-u^2)|^{-1}.\label{cor2}
\end{align}
At $\alpha=\alpha_{\min}=e^{3/2}/2$ and $|u|=\sqrt{3/2}$, the amplification reaches $A\rightarrow\infty$ \citep{Cordes17}. The actual physical optics gains should be finite, and the maximum value is
\begin{align}
A_{\max}&\sim\frac{a}{r_{\rm F}}=a\sqrt{\frac{2\pi d_{\rm os}}{\lambda d_{\rm ol}d_{\rm ls}}}\nonumber\\
&\simeq3.9~\nu_9^{1/2}a_{\rm AU,-3}d_{\rm os,Gpc}^{1/2}d_{\rm ol,Gpc}^{-1/2}d_{\rm ls,pc}^{-1/2},
\end{align}
where $r_{\rm F}=\sqrt{\lambda d_{\rm ol}d_{\rm ls}/2\pi d_{\rm os}}$ is the Fresnel scale, $a_{\rm AU,-3}=a/10^{-3}{\rm AU}$, $d_{\rm os,Gpc}=d_{\rm os}/{\rm Gpc}$, $d_{\rm ol,Gpc}=d_{\rm ol}/{\rm Gpc}$, and $d_{\rm ls,pc}=d_{\rm ls}/{\rm pc}$.
The focal frequency is defined as
\begin{align}
\nu_{\rm f}
&=\nu\left(\frac{\alpha}{\alpha_{\min}}\right)^{1/2}
=\frac{c}{a}\left(\frac{r_e{\rm DM}_l}{\pi \alpha_{\min}}\frac{d_{\rm ol}d_{\rm ls}}{d_{\rm os}}\right)^{1/2}\nonumber\\
&\simeq0.12~{\rm GHz}~{\rm DM}_{l,-8}^{1/2}a_{\rm AU,-3}^{-1}d_{\rm os,Gpc}^{-1/2}d_{\rm ol,Gpc}^{1/2}d_{\rm ls,pc}^{1/2},
\end{align}
where ${\rm DM}_{l,-8}={\rm DM}_l/(10^{-8}~{\rm pc~cm^{-3}})$.
The frequency below $\nu_{\rm f}$ will show ray crossings \citep{Cordes17}. 

According to Eq.(\ref{cor1}) and Eq.(\ref{cor2}), one can obtain the amplification as a function of the dimensionless frequency $\nu/\nu_{\rm f}$, as shown in Figure 5 of \citet{Cordes17}. There are two
caustic peaks in the amplified spectra with widths of $(1-10)\%$ of the observation frequency $\nu$, which are contributed by individual subimages by the Gaussian plasma lensing. The amplification is suppressed below the unit at the frequency $\nu\ll\nu_{\rm f}$ and asymptotes to the unity at $\nu\gg\nu_{\rm f}$ due to the caustic.

Similar to the scenario of gravitational lensing discussed in Section \ref{gravitational_lens}, if the time delays of subimaged bursts are smaller than their durations, the total image would be interfered, leading to the oscillating structures on the spectra. Due to the plasma dispersion along the line of sight, the arrival times of subimages should be chromatic, here we are only interested in the perturbations from the plasma lens, which add to delays from other plasma components (e.g., interstellar medium, intergalactic medium, etc.). The typical delay timescale of the plasma lensing is about \citep{Cordes17}
\begin{align}
\delta t\sim\frac{cr_e{\rm DM}_l}{2\pi\nu^2}\simeq4.1\times10^{-11}~{\rm s}~{\rm DM}_{l,-8}\nu_9^{-2}.
\end{align}
The phase difference of the rays from the two images is $\delta\phi=2\pi\nu\delta t$. A plasma lensing event with remarkable spectral modulation requires $\delta\phi\sim1$, $\nu\sim\nu_{\rm f}$ and $A_{\max}>1$, leading to
\begin{align}
&{\rm DM}_l\sim\frac{\nu}{cr_e}\simeq3.8\times10^{-8}~{\rm pc~cm^{-3}}\nu_9,\\
&a_{\rm AU,-3}^{-1}d_{\rm ls,pc}^{1/2}\sim4.1.
\end{align}
Therefore, a remarkable spectral modulation requires that the lensing object has an average electron number density of $n_e\sim 10~{\rm cm^{-3}}$ and a typical scale of $a\sim10^{-3}~{\rm AU}$ at a distance of $d_{\rm ls}\sim1~{\rm pc}$ from the FRB source. 

The variation timescale of the spectra depends on the changes in amplification through caustics due to the motions of source and observer. The effective transverse velocity combined into the motions of the source and the observer is $v_\perp=(d_{\rm ol}/d_{\rm os})v_{\rm s,\perp}+(d_{\rm ls}/d_{\rm os})v_{\rm o, \perp}$.
Using Eq.(\ref{cor1}) and Eq.(\ref{cor2}) and taking the derivatives, one has $\delta u'\sim(\delta A/A)A^2$ for $u\sim1$ when the amplification reaches the maximum value. Using $\delta u'\sim v_\perp t_{\rm cau}/a$, the timescale of a caustic crossing (also the variation timescale of the spectra) is estimated as
\begin{align}
t_{\rm cau}\sim\frac{a(\delta A/A)}{v_\perp A^2}\simeq15~{\rm s}~a_{\rm AU,-3}v_{\perp,7}^{-1}A_{1}^{-2}(\delta A/A).
\end{align}
Therefore, a plasma lensing can cause the burst-to-burst variation of the spectra during a much shorter time compared with that of the gravitational lensing. The variation timescale of the spectra of the two components of FRB 200428 might a relatively extreme amplification of $A\sim100$.

\section{Discussions and Conclusions}\label{summary}

Some FRBs appear significantly intrinsic narrow spectra in the telescope's bandwidth \citep{Pleunis21,Kumar21b,Zhou22}, which is an important clue to reveal the radiation mechanism and the environment of FRBs. 
In this work, we investigated the physical origin of the narrow spectra of FRBs from the perspectives of radiation mechanisms, coherent processes, radiative transfers, and interference processes, and the following conclusions are drawn:

1. Without considering the finite bandwidth of a telescope, the relative spectral bandwidth defined by FWHM or FWTM only depends on the intrinsic spectral shape. 
Some FRBs (e.g., FRB 20190711A, \citet{Kumar21b}) showed that their spectra must be intrinsically narrow $\Delta\nu/\nu_0\ll0.2$ for FWHM. This gives a strong constraint on the radiation mechanisms of FRBs because most radiation mechanisms with low-frequency spectral index $\alpha_l<3$ lead to wide spectra with $\Delta\nu/\nu_0>0.2$ for FWHM. 
In reality, the narrow telescope's bandwidth usually makes some observed bursts' spectra incomplete, and the distribution of the observed relative spectral bandwidths would be affected by the limited telescope's bandwidth.

2. An intrinsic narrow spectrum with $\Delta\omega/\omega_0\ll1$ ($\omega_0$ is the peak frequency and $\Delta\omega$ is the spectral bandwidth) implies that the electromagnetic wave is quasi-sinusoid with a typical frequency of $\omega\sim\omega_{0}$ in a short term and have a typical pulse duration of $T\sim4\pi/\Delta\omega$ in a long term. 
We generally discuss the spectral shapes and polarization distributions from the perspective of radiation mechanisms. For the radiation mechanisms involving the relativistic particle's perpendicular acceleration, the radiation features (including the spectrum and polarization) depend on the relation between the particle's deflection angle $\psi$ and the radiation beaming angle $1/\gamma$. The scenarios with $\gamma\psi\gg1$ and $\gamma\psi\ll1$ lead to different features of the spectrum and polarization.

3. For the scenario of $\gamma\psi\gg1$, the observer would see radiation from short segments of the particle's trajectory that are nearly parallel to the line of sight. Such a scenario is applicable to the curvature radiation and the traditional (large-pitch-angle) synchrotron radiation. The intrinsic spectra of these mechanisms are usually wide, and the intrinsic linear/circular polarization degree mainly depends on the angle between the viewing direction and the trajectory plane. 
If $\gamma\psi\ll1$, the particle's entire trajectory would be seen by the observer during a long term. 
In particular, for the small-pitch-angle synchrotron radiation by a single charged particle, the radiation is only emitted at a certain frequency within an extremely narrow band for a certain viewing direction, and the circular polarization is dominated. 
Furthermore, we discussed some astrophysical scenarios that might involve radiation processes with $\gamma\psi\gg1$ and $\gamma\psi\ll1$. In the magnetosphere of a neutron star, the radiation process with $\gamma\psi\gg1$ occurs in the inner magnetosphere and the corresponding radiation mechanism is the curvature radiation and the radiation process with $\gamma\psi\ll1$ occurs in the outer magnetosphere and the corresponding radiation mechanism is the small-pitch-angle synchrotron radiation. 
On the other hand, both scenarios with $\gamma\psi\gg1$ and $\gamma\psi\ll1$ might occur in the magnetized shocked medium, which corresponds to the traditional synchrotron radiation and the small-pitch-angle radiation, respectively. However, the generation of the small-pitch-angle radiation requires that the direction of the particles' injection is almost parallel to the field lines, which seems fine-tuning.

4. We find that coherent radiation processes can generate narrow spectra. For the bunching mechanisms (e.g., coherent curvature radiation), one possibility to generate a narrow spectrum is that the radiating bunches are quasi-periodically distributed. The quasi-periodic distribution of bunches might be due to the quasi-monochromatic Langmuir waves or quasi-periodic oscillating pair creation in the charge-starved region. For a train of bunches separated by a period of $1/\omega_m$, the total radiation will be coherently amplified at the peak frequencies of $\omega=2n\pi\omega_m$ with $n\in\mathbb{Z}^+$ that is independent of the radiation mechanism of the bunches. In particular, we notice that the bunches are not required to be positioned at exactly the same distance from each other. A small random relative phase of $\delta\phi_j\ll\pi$ can still lead to the coherent amplification at $\omega=2n\pi\omega_m$. However, once $\delta\phi_j\sim\pi$, the total radiation would become completely incoherent. 
On the other hand, the maser mechanism can naturally arise a narrow spectrum because a negative optical depth of $\tau_\nu$ causes an amplification by a factor of $\exp(|\tau_\nu|)$. An intermediate negative optical depth naturally causes the spectrum narrow.

5. The radiative transfer processes mainly include absorption and scattering. Based on the current observation, most processes seem not to significantly change the observed FRB spectra or directly arise a narrow spectral shape. The absorption processes can cause the radio bursts emitted during a short term to have the same low-frequency cutoff at the absorption frequency, which is inconsistent with the observations of some FRBs, e.g., the burst-to-burst spectral variation of two components of FRB 200428. Inverse Compton scattering suppresses the specific intensity in all bands but can not change the spectral shape. Induced Compton scattering redistributes the incident photons toward low frequencies, and the photons are eventually absorbed by the absorption processes, leading to the same issue as the absorption. The nonlinear scattering depends on the radiation intensity and frequency and makes the low-frequency cutoff proportional to the intensity, which is worthy to test the nonlinear effect by the observations of FRB repeaters.

6. We discussed some interference processes including scintillation, gravitational lensing, and plasma lensing. If the scintillation significantly changes the observed shape, the scintillation bandwidth is required to be $\Delta\nu_{\rm sci}\gtrsim100~{\rm MHz}$, which is much larger than the observed narrow-bandwidth scintillation with $\Delta\nu_{\rm sci}\sim1~{\rm MHz}$. Thus, involving another plasma screen is necessary to modulate the spectra at a bandwidth of $\gtrsim100~{\rm MHz}$. The scintillation with $\Delta\nu_{\rm sci}\gtrsim100~{\rm MHz}$ requires that the corresponding plasma screen is within a distance of $\sim10^{15}~{\rm cm}$ from the FRB source and the plasma medium is intermediately dense and turbulent. The spectral modulations of gravitational lensing and plasma lensing can modulate the FRB spectra when the time delay of subimaged bursts is about $\delta t\sim10^{-10}~{\rm s}$ for GHz wave. For gravitational lensing, such a condition requires that the lensing object has a mass of $\sim10^{-5}M_\odot$, i.e., a planet-like object. However, since the gravitational lensing event is most likely one-time but the spectrum variation is always present during the several-year active period of an FRB repeater, such a scenario could be ruled out. Besides, the gravitational lensing cannot explain the remarkable burst-to-burst variation of the spectra in a time shorter than the typical lensing timescale.
The delay time of $\delta t\sim10^{-10}~{\rm s}$ requires the plasma lens has an average electron number density of $n_e\sim 10~{\rm cm^{-3}}$ and a typical scale of $a\sim10^{-3}~{\rm AU}$ at a distance of $d_{\rm ls}\sim1~{\rm pc}$ from the FRB source. These conditions are moderate to explain the observed FRB spectra. Meanwhile, the typical lensing time scale of decades of seconds can cause the spectrum variation during a short time, although the extremely short spectrum variation as shown in FRB 200428 might be due to a relatively extreme amplification.

\acknowledgments

We thank the anonymous referee for the helpful comments and suggestions.
We also thank Bing Zhang and Yue Wu for reading the initial manuscript and for their helpful comments and acknowledge the discussions with Yi Feng, Jin-Lin Han, Kejia Lee, Ze-Nan Liu, Yuanhong Qu, Wei-Yang Wang, and Yong-Kun Zhang.
This work is supported by the National Natural Science Foundation of China grant No.12003028 and the National SKA Program of China (2022SKA0130100). 

\appendix

\section{Radiations by a single charged particles}\label{mechanism_appendix}

In this appendix, we generally discuss the features of the spectra and polarizations of the radiation by charged accelerated particles, which are applicable to most radiation mechanisms in various astrophysical scenarios.
We consider that a particle with a charge $q$ moves on a trajectory $\vec{r}_0(t')$ with velocity $\vec{\beta}(t')c$ and acceleration $\dot{\vec{\beta}}(t')c$, where $t'$ is the retarded time. The line-of-sight direction is $\vec{n}$ and the distance between the particle and the observer is $R$. The radiation field at $(\vec{r},t)$ is given by \citep[e.g.,][]{Rybicki86}
\begin{align}
\vec{E}(\vec{r},t)&=\frac{q}{c}\left[\frac{\vec{n}}{\kappa^3R}\times\{(\vec{n}-\vec{\beta})\times\dot{\vec{\beta}}\}\right]_{\rm rec},\nonumber\\
\vec{B}(\vec{r},t)&=\left[\vec{n}\times\vec{E}\right]_{\rm rec},
\end{align}
where $\kappa\equiv1-\vec{n}\cdot\vec{\beta}$, the quantities in the square bracket, $\left[...\right]_{\rm rec}$, are evaluated at the retarded time $t'$. 
The radiation energy per unit frequency interval per unit solid angle is
\begin{align}
\mathcal{E}_\omega&\equiv\frac{dW}{d\omega d\Omega}=\frac{q^2}{4\pi^2c}\left|\int_{-\infty}^{\infty}\left[\kappa^{-3}\vec{n}\times\{(\vec{n}-\vec{\beta})\times\dot{\vec{\beta}}\}\right]_{\rm rec}e^{i\omega t}dt\right|^2\label{eq1}\\
&=\frac{q^2\omega^2}{4\pi^2c}\left|\int \vec{n}\times(\vec{n}\times\vec{\beta})\exp\left[i\omega(t'-\vec{n}\cdot\vec{r}_0(t')/c)\right]dt'\right|^2\\
&=\frac{e^2\omega^2}{4\pi^2c}\left|-\vec{\epsilon_\parallel} A_\parallel(\omega)+\vec{\epsilon_\perp}A_\perp(\omega)\right|^2,
\end{align}
where $A_\parallel$ and $A_\perp$ are two orthogonal components perpendicular to the line of sight. 
The linear polarization degree is
\begin{align}
\pi_L=
\left|\frac{\left[(A_\parallel A_\parallel^*-A_\perp A_\perp^*)^2+(A_\parallel A_\perp^*+A_\perp A_\parallel^*)^2\right]^{1/2}}{A_\parallel A_\parallel^*+A_\perp A_\perp^*}\right|,\label{pil}
\end{align}
and the circular polarization degree is
\begin{align}
\pi_V=\left|\frac{1}{i}\frac{A_\parallel A_\perp^*-A_\perp A_\parallel^*}{A_\parallel A_\parallel^*+A_\perp A_\perp^*}\right|.\label{piv}
\end{align}
It should be noted that the $\mathcal{E}_\omega$ corresponds to the total radiation energy in an entire pulse.
If the radiation pulse repeats on average time $T$, the above radiation energy could be converted to the radiation power \citep{Rybicki86}
\begin{align}
\mathcal{P}_\omega\equiv\frac{1}{T}\frac{dW}{d\omega d\Omega}=\frac{\mathcal{E}_\omega}{T}.
\end{align}

The radiation spectrum by a single charged particle with the perpendicular acceleration depends on the relation between the particle's deflection angle $\psi$ and the radiation beaming angle $\sim1/\gamma$ \citep{Landau75}, as shown in Figure \ref{emission_trajectory}, which will be summarized in the following subsections.

\subsection{Deflection angle larger than radiation beaming angle}\label{mechanism1a}

In the scenario with $\gamma\psi\gg1$, the radiation is equivalent to the radiation by the particle moving instantaneously at constant speed on an appropriate circular path \citep{Jackson98}, as shown in the panel (a) of Figure \ref{emission_trajectory}.
We consider that the acceleration curvature radius is $\rho$, the angle between the line of sight and the trajectory plane is $\theta$, and the radiation angular frequency is $\omega$.
The radiation energy per unit frequency interval per unit solid angle is given by the above equation with \citep{Jackson98}
\begin{align}
A_\parallel(\omega)&=\frac{2i}{\sqrt{3}}\frac{\rho}{c}\left(\frac{1}{\gamma^2}+\theta^2\right)K_{2/3}(\xi),\label{CR_A1}\\
A_\perp(\omega)&=\frac{2}{\sqrt{3}}\frac{\rho\theta}{c}\left(\frac{1}{\gamma^2}+\theta^2\right)^{1/2}K_{1/3}(\xi),\label{CR_A2} 
\end{align}
and
\begin{align}
\xi\equiv\frac{1}{2}\hat\omega\left(1+\gamma^2\theta^2\right)^{3/2}~~~\text{with}~~~\hat\omega\equiv\frac{\omega}{\omega_c},
\end{align}
where $\omega_c=3\gamma^3c/2\rho$ is the typical radiation frequency.
The radiation energy per unit frequency interval per unit solid angle is
\begin{align}
\mathcal{E}_\omega=\frac{3e^2}{4\pi^2c}\gamma^2\hat\omega^2(1+\gamma^2\theta^2)^2\left[K_{2/3}^2(\xi)+\frac{1}{1/\gamma^2\theta^2+1}K_{1/3}^2(\xi)\right].\label{CRspec}
\end{align}

\begin{figure}
    \centering
    \includegraphics[width = 0.5\linewidth, trim = 0 0 0 0, clip]{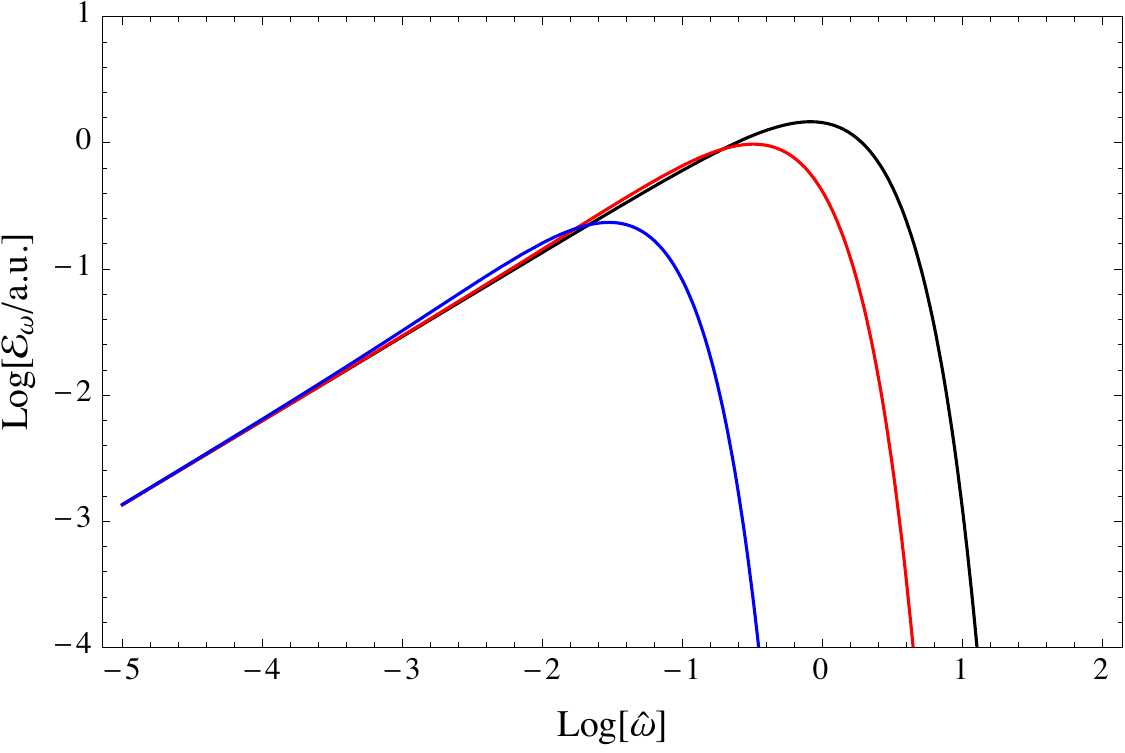}
    \caption{The spectrum given by Eq.(\ref{CRspec0}) that is applicable for the scenario with the particle's deflection angle larger than the radiation beaming angle, $\gamma\psi\gg1$. The black, red, and blue lines correspond to $\gamma\theta=0.1,1~{\rm and}~3$, respectively. The unit of $\mathcal{E}_\omega$ is arbitrary.}\label{CR_spectrum} 
\end{figure}

\begin{figure}
    \centering
    \includegraphics[width = 0.45\linewidth, trim = 0 0 0 0, clip]{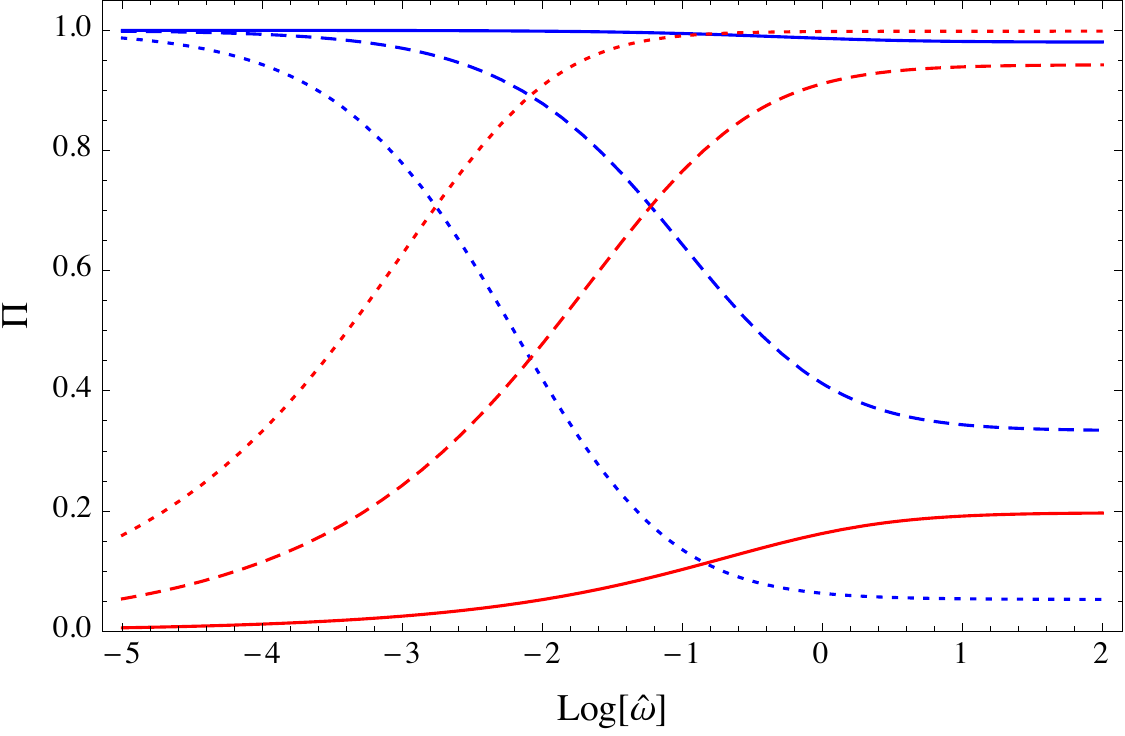}
    \includegraphics[width = 0.45\linewidth, trim = 0 0 0 0, clip]{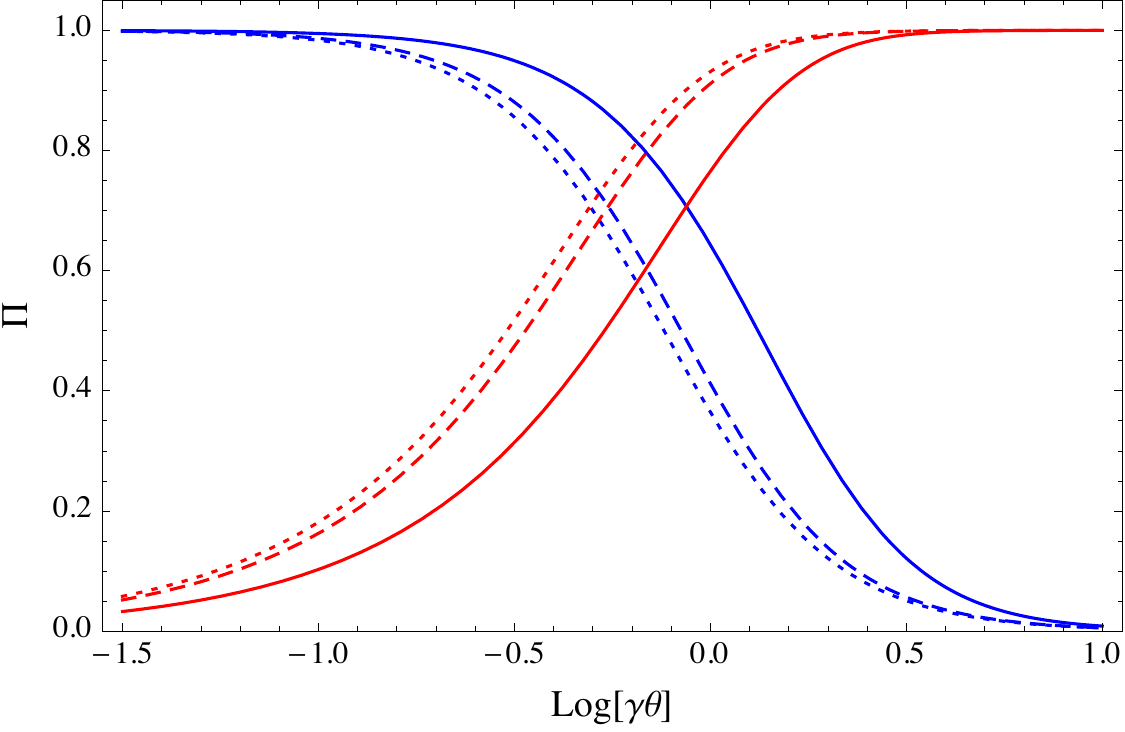}
    \caption{The relations between the polarization degree $\Pi$ and the dimensionless frequency $\hat\omega$ and the viewing angle $\gamma\theta$ for a single radiating particle with $\gamma\psi\gg1$. The blue and red lines correspond to the linear and circular polarization degrees, respectively. The top panel shows the polarization degree $\Pi$ as the function of the dimensionless frequency $\hat\omega$. The solid, dashed, and dotted lines correspond to $\gamma\theta=0.1,1~{\rm and}~3$, respectively. The bottom panel shows the polarization degree $\Pi$ as the function of the viewing angle $\gamma\theta$. The solid, dashed, and dotted lines correspond to $\hat\omega=0.1,1~{\rm and}~3$, respectively.}\label{pol_freq} 
\end{figure}

The spectrum $\mathcal{E}_\omega$ of a single radiating particle satisfies the power-law distribution with $\mathcal{E}_\omega\propto\hat\omega^{2/3}$ at the low frequency and appears an exponential decay at the high frequency, which is intrinsically wide \citep[e.g.,][]{Jackson98,Yang18}, $\Delta\omega/\omega_0\sim1$ (see Section \ref{specband}), as shown in Figure \ref{CR_spectrum}. Meanwhile, the larger the viewing angle $\gamma\theta$, the lower the cutoff frequency.
The spectrum of the radiation by multiple particles is also usually wide, which has been discussed in detail in \citep{Yang18,Yang23} and we will not repeat it here. 
According to Eq.(\ref{pil}), Eq.(\ref{piv}), Eq.(\ref{CR_A1}) and Eq.(\ref{CR_A2}), the linear polarization degree is
\begin{align}
\pi_L=\left|\frac{K_{2/3}^2(\xi)-1/(1/\gamma^2\theta^2+1)K_{1/3}^2(\xi)}{K_{2/3}^2(\xi)+1/(1/\gamma^2\theta^2+1)K_{1/3}^2(\xi)}\right|,\label{CR_pil}
\end{align}
and the circular polarization degree is
\begin{align}
\pi_V=\left|\frac{2/(1/\gamma^2\theta^2+1)^{1/2}K_{2/3}(\xi)K_{1/3}(\xi)}{K_{2/3}^2(\xi)+1/(1/\gamma^2\theta^2+1)K_{1/3}^2(\xi)}\right|.\label{CR_piv}
\end{align}
Similar to the spectrum given by $\mathcal{E}_\omega$, both $\pi_L$ and $\pi_V$ are also the functions as the variables $(\hat\omega,\gamma\theta)$.
In Figure \ref{pol_freq}, we plot the linear and circular polarization degrees $\Pi_i$ with $i=L~{\rm and}~V$ as the functions of the dimensionless frequency $\hat\omega$ and the viewing angle $\gamma\theta$, respectively. For a certain viewing angle $\gamma\theta$, the higher the frequency $\hat\omega$, the lower (higher) the linear (circular) polarization degree. For a certain frequency $\hat\omega$, the larger the viewing angle $\gamma\theta$, the lower (higher) the linear (circular) polarization degree. Thus, the high circular polarization degree should be attributed to the off-beam observation \citep{Wang22c,Liu22,Qu23b}.

The polarization measurement at least requires that burst flux larger than the telescope's threshold.
For the scenario with $\gamma\psi\gg1$, according to Eq.(\ref{CRspec0}) and using the property of Bessel function, $K_{\nu}(x)\sim\sqrt{\pi/2x}\exp(-x)$ for $x\gg1$, the radiation energy falls off in angle approximately as
\begin{align}
\mathcal{E}_\omega\sim\mathcal{E}_{\omega,0}\exp\left(-\hat\omega\gamma^3\theta^3\right),
\end{align}
where $\mathcal{E}_{\omega,0}\equiv\mathcal{E}_\omega(\theta=0)$ and $\mathcal{E}_\omega\propto K_{\nu}^2(\xi)$ is used. 
We consider that only the burst with radiation energy
\begin{align}
\mathcal{E}_\omega>\eta_c^{-1}\mathcal{E}_{\omega,0}~~~\text{with}~\eta_c\geqslant 1
\end{align} 
could be observed due to the constraint of a detector's sensitivity, 
leading to
\begin{align}
\gamma\theta<\gamma\theta_{\rm th}\equiv\left(\frac{1}{\hat\omega}\ln\eta_c\right)^{1/3}.\label{gamtheth}
\end{align}
Only the bursts with viewing angle $\theta<\theta_{\rm th}$ are observable. Notice that the distribution of the intrinsic burst energies is neglected, and here we are mainly interested in the suppression effect by the viewing direction.

\begin{figure*}
    \centering
    \includegraphics[width = 0.3\linewidth, trim = 0 0 0 0, clip]{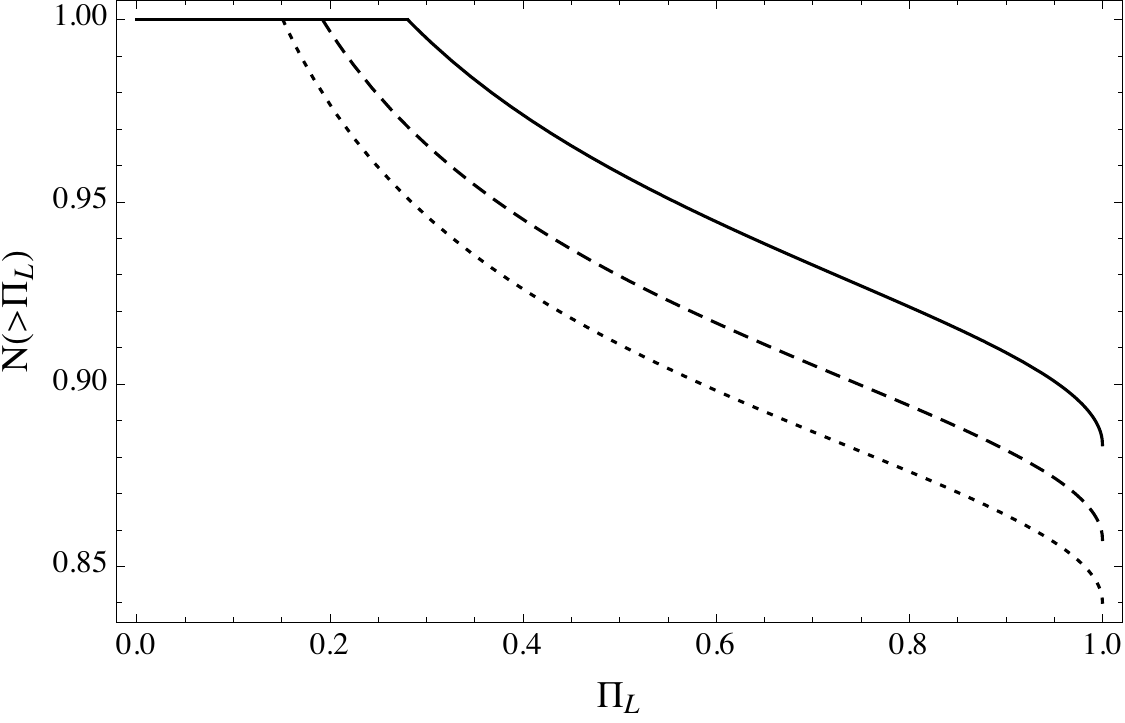}
    \includegraphics[width = 0.3\linewidth, trim = 0 0 0 0, clip]{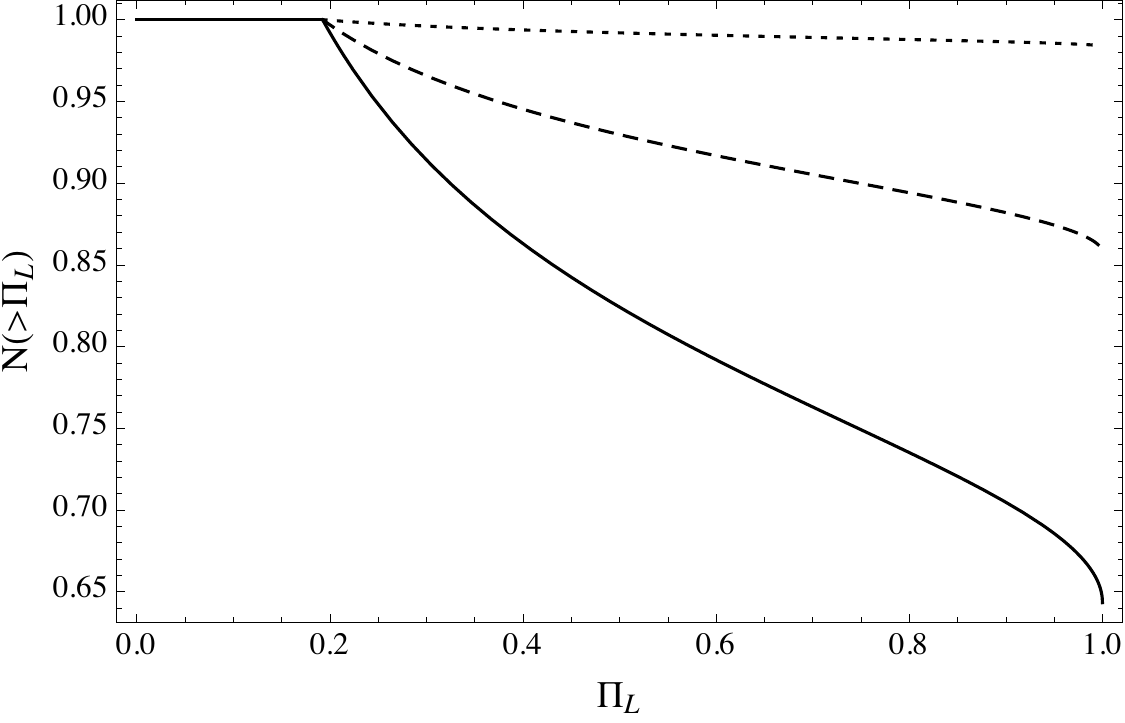}
    \includegraphics[width = 0.3\linewidth, trim = 0 0 0 0, clip]{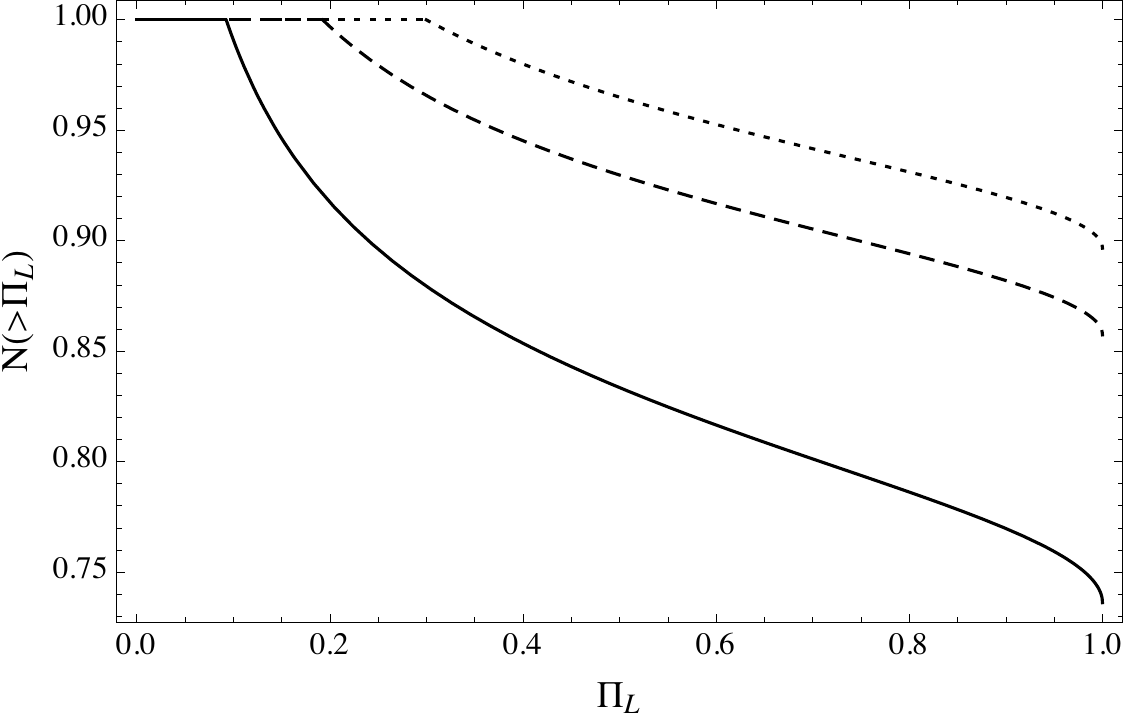}
    \includegraphics[width = 0.3\linewidth, trim = 0 0 0 0, clip]{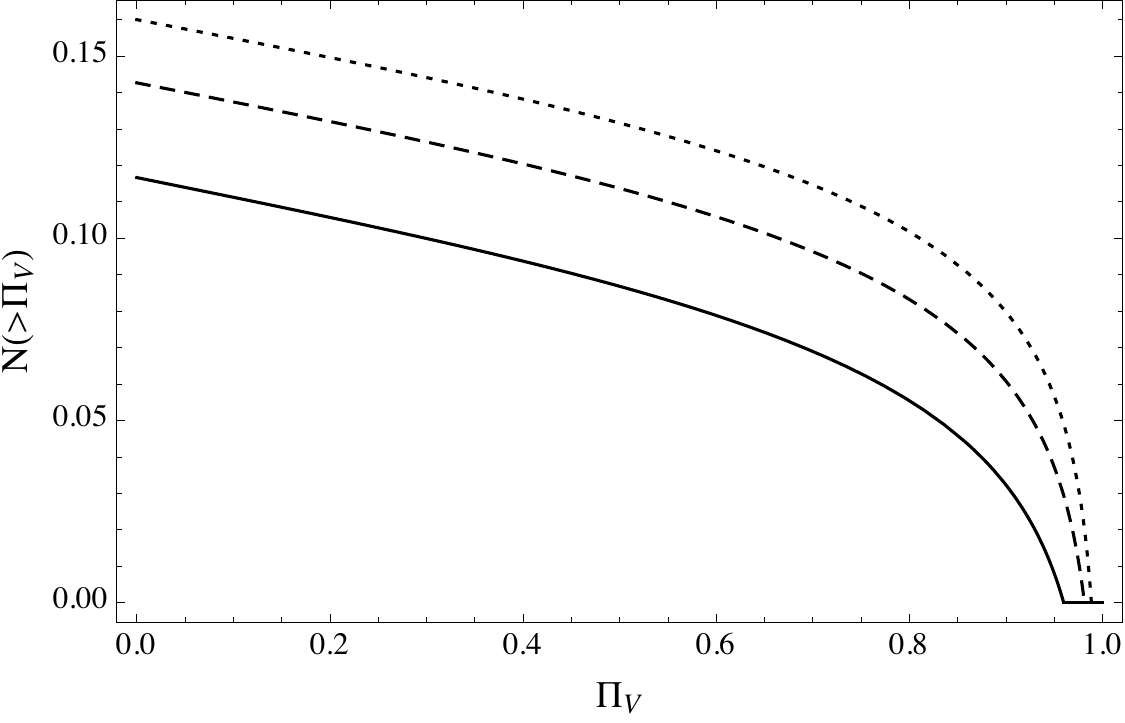}
    \includegraphics[width = 0.3\linewidth, trim = 0 0 0 0, clip]{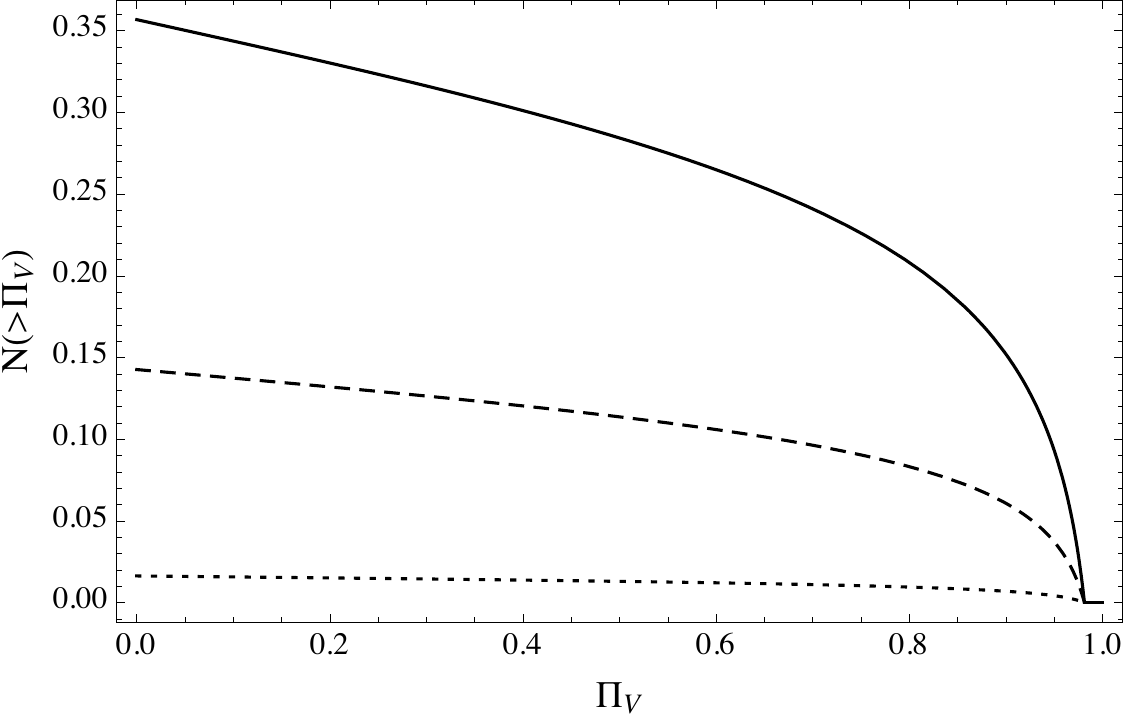}
    \includegraphics[width = 0.3\linewidth, trim = 0 0 0 0, clip]{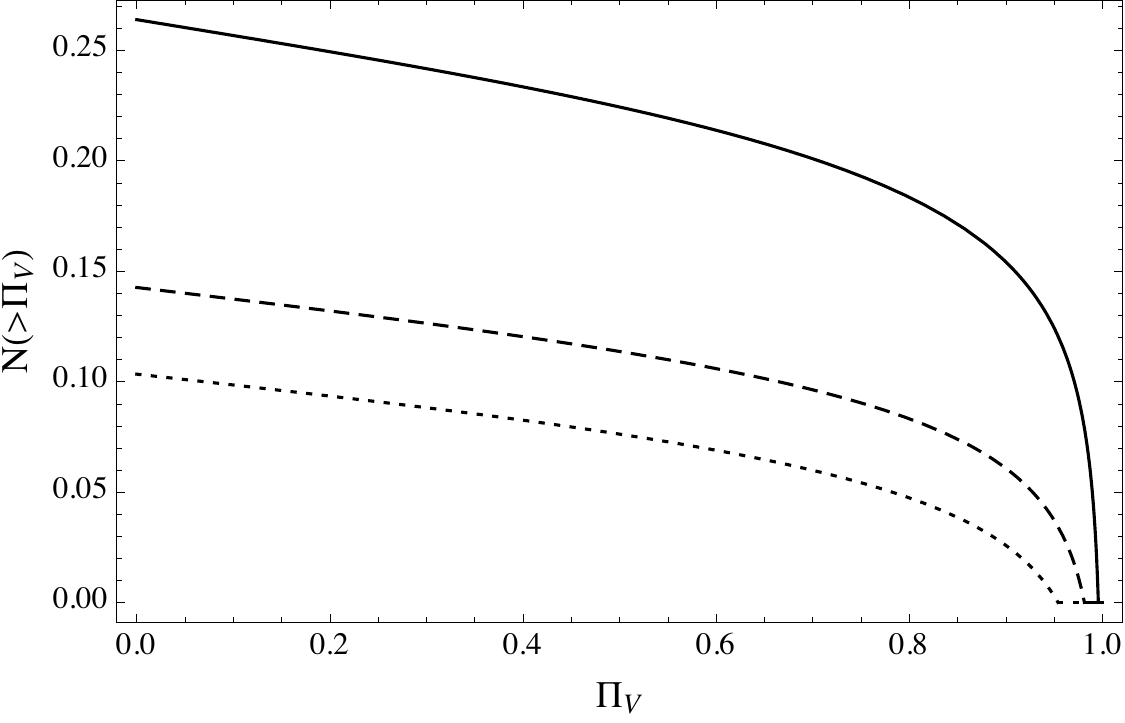}
    \caption{The cumulative distribution of the linear (top panels) and circular (bottom panels) polarization degrees for the radiation by multiple particles with $\gamma\psi\gg1$. Notice that the cumulative distribution of the circular polarization degrees is $N(>\Pi_V=0)=1$ at $\Pi_V=0$ and decreases significantly once $\Pi_V>0$. The component of $N(>\Pi_V=0)=1$ at $\Pi_V=0$ is not shown in the bottom panels.
    The distribution function of the viewing direction is assumed to satisfy Eq.(\ref{CR_dist1}). In the left panels: the solid, dashed, and dotted lines correspond to $\eta_c=10,100~{\rm and}~1000$, respectively for $\hat\omega=1$ and $\gamma\Theta_j=10$. In the middle panels: the solid, dashed, and dotted lines correspond to $\gamma\Theta_j=3,10~{\rm and}~100$, respectively for $\hat\omega=1$ and $\eta_c=100$. In the right panels: the solid, dashed, and dotted lines correspond to $\hat\omega=0.1,1~{\rm and}~3$, respectively for $\gamma\Theta_j=10$ and $\eta_c=100$.}\label{cum_dist1}
\end{figure*}

We consider that multiple radiating particles are uniformly distributed in a fan beam with an opening angle $\Theta_j$, and the viewing angle is $\Theta$.
According to Eq.(\ref{CR_pil}) and Eq.(\ref{CR_piv}), the polarization is 100\% linear when the viewing direction is on the trajectory plane, and most radiation energy is emitted near the trajectory plane. The larger the viewing angle, the higher the circular polarization degree. Since the circular polarizations on the different sides of the trajectory plane are opposite, in the particles' beam center the coherent sum of the circular polarizations will be canceled, leading to the linear polarization dominated. A detailed analysis was also discussed in \citet{Wang22b} and \citet{Liu22}.
Therefore, the linear polarization degree could be approximately given by
\begin{align}
\Pi_L\simeq\left\{
\begin{aligned}
&1,&&\text{for}~\Theta\leqslant\Theta_j,\\
&\pi_L(\Theta-\Theta_j),&&\text{for}~\Theta>\Theta_j, 
\end{aligned}
\right.
\end{align}
and the circular polarization degree could be approximately given by
\begin{align}
\Pi_V\simeq\left\{
\begin{aligned}
&0,&&\text{for}~\Theta\leqslant\Theta_j,\\
&\pi_V(\Theta-\Theta_j),&&\text{for}~\Theta>\Theta_j, 
\end{aligned}
\right.
\end{align}
Since the view direction related to the trajectory plane is random, the number of the  bursts emitting within $(\Theta,\Theta+d\Theta)$ is
\footnote{Notice that the distribution of the viewing direction should be not $N(\Theta)d\Theta=\sin\Theta d\Theta$ in this scenario, because the viewing direction is related to the trajectory plane of the accelerated particle, see Figure 14.9 and Section 14 in \citet{Jackson98} for a detailed discussion.}
\begin{align}
N(\Theta)d\Theta=\frac{1}{(\Theta_j+\theta_{\rm th})} d\Theta,\label{CR_dist1}
\end{align}
where $\theta_{\rm th}$ is given by Eq.(\ref{gamtheth}), and $\Theta_j+\theta_{\rm th}$ corresponds to the threshold angle above which the observed flux would be less than the telescope's flux threshold.
Therefore, the cumulative distribution of the linear and circular polarization degrees are
\begin{align}
N(>\Pi_L)&=\frac{\gamma\Theta_j+(\gamma\theta)(\Pi_L)}{\gamma\Theta_j+\gamma\theta_{\rm th}},\label{CR_npil}\\
N(>\Pi_V)&=\frac{\gamma\theta_{\rm th}-(\gamma\theta)(\Pi_V)}{(\gamma\Theta_j+\gamma\theta_{\rm th})},\label{CR_npiv}
\end{align}
for $(\gamma\theta)(\Pi_i)\leqslant\gamma\theta_{\rm th}$ with $i=L,V$, respectively,
where $(\gamma\theta)(\Pi_i)$ is the inverse function of $\pi_i(\gamma\theta)$ given by Eq.(\ref{CR_pil}) and Eq.(\ref{CR_piv}).

In Figure \ref{cum_dist1}, we plot the cumulative distributions of the linear and circular polarization degrees according to Eq.(\ref{CR_npil}) and Eq.(\ref{CR_npiv}). The cumulative distributions of the polarization degrees depend on the telescope's flux threshold $\eta_c$, the particles' beaming angle $\gamma\Theta_j$, and observed frequency $\hat\omega$. We can see that: (1) The higher the telescope's sensitivity (i.e., a large value of $\eta_c$), the lower the number fraction between the linearly and circularly polarized bursts. The reason is that most high circularly polarized bursts have relatively low fluxes due to large values of $\gamma\theta$. (2) The larger the particles' beaming angle, the higher the number fraction between the linearly and circularly polarized bursts. If $\gamma\Theta_j\gg1$, most bursts would have $\Pi_L\sim1$ and $\Pi_V\sim0$. A moderate number fraction between linearly polarized bursts and circularly polarized bursts as shown in FRB 20201124A \citep{Xu21,Jiang22} requires that $\gamma\Theta_j\sim1$. (3) The higher the observed frequency, the higher the number fraction between the linearly and circularly polarized bursts. The reason is that the threshold viewing angle $\gamma\theta_{\rm th}$ is significantly suppressed at the high frequency (see Eq.(\ref{gamtheth})), leading to a larger relative number of the bursts from the particle beaming angle $\Theta_j$.

\subsection{Deflection angle smaller than radiation beaming angle}\label{mechanism2a}

In the scenario with $\gamma\psi\ll1$, the particle with a charge $q$ moves along the line of sight with an almost constant velocity $\vec{\beta}$ but with a varying acceleration $\dot{\vec{\beta}}$, as shown in the panel (b) of Figure \ref{emission_trajectory}. 
The radiation energy per unit frequency interval per unit solid angle at the line-of-sight direction $\vec{n}$ could be written as \citep[][also see Appendix \ref{mechanism_appendix}]{Landau75}
\begin{align}
\mathcal{E}_\omega
=\frac{q^2}{4\pi^2c}\left(\frac{\omega}{\tilde\omega}\right)^4\left|\vec{n}\times\left[(\vec{n}-\vec{\beta})\times\dot{\vec{\beta}}_{\tilde\omega}\right]\right|^2\label{SPA_spec}
\end{align}
with
\begin{align}
\dot{\vec{\beta}}_{\tilde\omega}\equiv\int_{-\infty}^{\infty}\dot{\vec{\beta}}e^{i\tilde\omega t'}dt'~~~\text{and}~~~\tilde\omega\equiv(1-\vec{n}\cdot\vec{\beta})\omega.
\end{align}
where $t'$ is the retarded time.
In the ultrarelativistic case, the longitudinal acceleration is smaller than the transverse acceleration, $\dot{\vec{\beta_\parallel}}/\dot{\vec{\beta_\perp}}\sim1/\gamma^2\ll1$. Thus, $\dot{\vec{\beta}}$ and $\vec{\beta}$ are approximately perpendicular to each other, $\dot{\vec{\beta}}\perp\vec{\beta}$. Since both $\vec{n}$ and $\vec{\beta}$ are approximately constant in the above equation, the properties of the spectrum and polarization are mainly determined by the acceleration $\dot{\vec{\beta}}$.

First, we discuss the general properties of the polarization in the scenario with $\gamma\psi\ll1$. Since both $\vec{n}$ and $\vec{\beta}$ are approximately constant in Eq.(\ref{SPA_spec}), the spectral properties are mainly determined by the acceleration $\dot{\vec{\beta}}$.
Choosing a coordinate system $S$ with $z$-direction pointing toward the observer and with the particle velocity on the $y-z$ plane, thus $\vec{n}=(0,0,1)$ and $\vec{\beta}=(0,\sin\theta,\cos\theta)$. In the coordinate system $S'$ with $z'$-direction pointing toward the particle velocity $\vec{\beta}$ and $x'$-axis parallel with $x$-axis, the acceleration could be written as $\dot{\vec{\beta}}'=(b\cos\phi,b\sin\phi,0)$, where $b=|\dot{\vec{\beta}}'|$ and $\phi$ is the azimuth angle of $\dot{\vec{\beta}}'$ in the $x'-y'$ plane that is perpendicular to $z'$-direction. Thus, the acceleration in the $S$ coordinate system is $\dot{\vec{\beta}}=(b\cos\phi,b\sin\phi\cos\theta,-b\sin\phi\sin\theta)$. According to Eq.(\ref{SPA_spec}), the radiation polarization property is determined by
\begin{align}
\vec{n}\times\left[(\vec{n}-\vec{\beta})\times\dot{\vec{\beta}}\right]=\left[-b\cos\phi(1-\cos\theta),b\sin\phi(1-\cos\theta),0\right],
\end{align}
leading to
\begin{align}
A_\parallel(\tilde\omega)&\propto-\int b(t)\cos\phi(t)e^{i\tilde\omega t}dt,\\
A_\perp(\tilde\omega)&\propto\int b(t)\sin\phi(t)e^{i\tilde\omega t}dt.
\end{align}
Based on Eq.(\ref{pil}) and Eq.(\ref{piv}), we can easily prove that: (1) If the acceleration is always on a straight line perpendicular to $\vec{\beta}$, i.e., $\phi={\rm const.}$, the polarization is fully linear with $\pi_L=1$; (2) If the acceleration rotates with a constant angular velocity $\Omega$ on the plane perpendicular to $\vec{\beta}$, i.e., $\phi(t)=\Omega t$ and $b(t)={\rm const}$, the polarization is fully circular with $\pi_V=1$.

In order to obtain the accurate spectrum and polarization, a simpler and more intuitive processing method is to calculate the radiation in the particle comoving frame $K'$ with velocity $\beta_\parallel=\beta\cos\psi$ related to the observer frame $K$, then transfer the radiation to the $K$ frame via Lorentz and Doppler transformations. 
In the $K'$ frame, the particle moves with the velocity 
\begin{align}
\beta'\simeq\frac{\beta_\perp}{\gamma(1-\beta_\parallel^2)}=\frac{(\gamma^2-1)^{1/2}\sin\psi}{\cos^2\psi+\gamma^2\sin^2\psi}\simeq\gamma\psi\ll1.
\end{align}
Thus, the particle in the $K'$ frame is non-relativistic for $\gamma\psi\ll1$. 
In many astrophysical scenarios, the perpendicular acceleration of a charged particle is usually attributed to the Lorentz force by magnetic fields, meanwhile, the intrinsic variation timescale of the magnetic field is longer than $\Delta t_{\rm acc}$.
In this case, the radiation in the $K'$ frame is cyclotron-like. We consider that in the $K'$ frame the acceleration curvature radius is $\rho'$, and the angle between the line of sight and the trajectory plane is $\theta'$. 
We define 
\begin{align}
\zeta\equiv m\beta'\sin\theta'
\end{align}
with $m$ as the harmonic number, then in the $K'$ frame, the radiation power per unit solid angle in the $m$-th harmonic is \citep{Landau75,Jackson98}
\begin{align}
\frac{dP_{m}'}{d\Omega}=\frac{e^2\omega_0^4 m^2}{8\pi^3 c}\left|-\vec{\epsilon_\parallel} A'_\parallel(\omega)+\vec{\epsilon_\perp}A'_\perp(\omega)\right|^2
\end{align}
with
\begin{align}
A'_\parallel(\omega)&=\frac{2\pi i\rho'}{c}\frac{dJ_m(\zeta)}{d\zeta},\label{SPA_A1}\\
A'_\perp(\omega)&=\frac{2\pi \rho'}{c}\frac{\cot\theta'}{\beta'}J_m(\zeta),\label{SPA_A2}
\end{align}
where the fundamental frequency is 
\begin{align}
\omega_0=\frac{\beta'c}{\rho'}.\label{freq2}
\end{align}
In particular, if the gyration motion is caused by the Lorentz force of the magnetic field, one has $\omega_0=\omega_B'=eB/m_ec$, where $\omega_B'$ is the cyclotron frequency in the $K'$ frame.
According to Eq.(\ref{pil}), Eq.(\ref{piv}), Eq.(\ref{SPA_A1}) and Eq.(\ref{SPA_A2}), the linear polarization degree is
\begin{align}
\pi_L=\left|\frac{[dJ_m(\zeta)/d\zeta]^2-(\cot^2\theta'/\beta'^2)J_m^2(\zeta)}{[dJ_m(\zeta)/d\zeta]^2+(\cot^2\theta'/\beta'^2)J_m^2(\zeta)}\right|,
\end{align}
and the circular polarization degree is 
\begin{align}
\pi_V=\left|\frac{2[dJ_m(\zeta)/d\zeta](\cot\theta'/\beta')J_m(\zeta)}{[dJ_m(\zeta)/d\zeta]^2+(\cot^2\theta'/\beta'^2)J_m^2(\zeta)}\right|.
\end{align}
Due to $\beta'\ll1$, the emission but the fundamental frequency $m=1$ can be neglected, leading to an extremely narrow spectrum. Using the properties of Bessel function $J_m(x)\sim[1/\Gamma(m+1)](x/2)^m$ for $0<x\ll(m+1)^{1/2}$, the radiation power reduces to
\begin{align}
\mathcal{P}'_\omega\equiv\frac{dP_{m}'}{d\omega d\Omega}=\frac{e^2\omega_0^2\beta'^2}{8\pi c}(1+\cos^2\theta')\delta(\omega'-\omega_0).
\end{align}
The linear polarization degree is
\begin{align}
\pi_L'=\left|\frac{1-\cos^2\theta'}{1+\cos^2\theta'}\right|,\label{SPA_pil}
\end{align}
and the linear polarization degree is
\begin{align}
\pi_V'=\left|\frac{2\cos\theta'}{1+\cos^2\theta'}\right|.\label{SPA_piv}
\end{align}
Using the following transformations
\begin{align}
&\cos\theta'=\frac{\cos\theta-\beta_\parallel}{1-\beta_\parallel\cos\theta}\simeq\frac{1-\gamma^2\theta^2}{1+\gamma^2\theta^2},\label{costhe}\\
&\omega'=\omega\frac{1-\beta_\parallel\cos\theta}{(1-\beta_\parallel^2)^{1/2}}\simeq\frac{\omega}{2\gamma}(1+\gamma^2\theta^2),\\
&\mathcal{P}_\omega=\mathcal{P}_\omega'\frac{(1-\beta_\parallel^2)^{3/2}}{(1-\beta_\parallel\cos\theta)^3}\simeq\mathcal{P}_\omega'\frac{8\gamma^3}{(1+\gamma^2\theta^2)^3},\label{PomegaT}
\end{align}
the received radiation power is
\begin{align}
\mathcal{P}_\omega=\frac{e^2\gamma^2\psi^2\omega^4}{4\pi c\omega_0^2}\left(1-\frac{\omega}{\gamma\omega_0}+\frac{\omega^2}{2\gamma^2\omega_0^2}\right)\delta\left(\omega-\frac{2\gamma\omega_0}{1+\gamma^2\theta^2}\right).
\end{align}
Note that the radiation power $\mathcal{P}_\omega$ in Eq.(\ref{PomegaT}) is emphasized to be the received specific power in the $K$ frame, and a factor of $\gamma^3(1+\beta\cos\theta')^3$ should be corrected from the emitted specific power, see Section 4.8 of \citet{Rybicki86} and involve $d\omega=\gamma(1+\beta\cos\theta')d\omega'$ to calculate the specific radiation power.
We define 
\begin{align}
\bar\omega\equiv\frac{\omega}{\gamma\omega_0},\label{freq3}
\end{align}
then the received radiation power can be rewritten as
\begin{align}
\mathcal{P}_\omega=\frac{e^2\gamma^5\psi^2\omega_0}{4\pi c}\bar\omega^4\left(1-\bar\omega+\frac{1}{2}\bar\omega^2\right)\delta\left(\bar\omega-\frac{2}{1+\gamma^2\theta^2}\right),\label{P_omega}
\end{align}
and the radiation only occurs at the direction $\theta$ with 
\begin{align}
\gamma\theta=\left(\frac{2}{\bar\omega}-1\right)^{1/2}.\label{omega_theta}
\end{align}
Notice that the particle's deflection angle $\psi$ only affects the normalized radiation power but not the typical frequency and the spectral shape.
According to Eq.(\ref{P_omega}), for a certain viewing direction $\gamma\theta$, the emission is only at the frequency $\bar\omega=2/(1+\gamma^2\theta^2)$. Thus, the radiation spectrum of a single particle is extremely narrow. Since most radiation energy is emitted with the direction satisfying $\gamma\theta\lesssim1$, the typical radiation frequency is $\bar\omega\sim{\rm a~few}$ (corresponding to $\omega\sim\gamma\omega_0$), which is consistent with the above result estimated by Eq.(\ref{freq10}).

\begin{figure}
    \centering
    \includegraphics[width = 0.45\linewidth, trim = 0 0 0 0, clip]{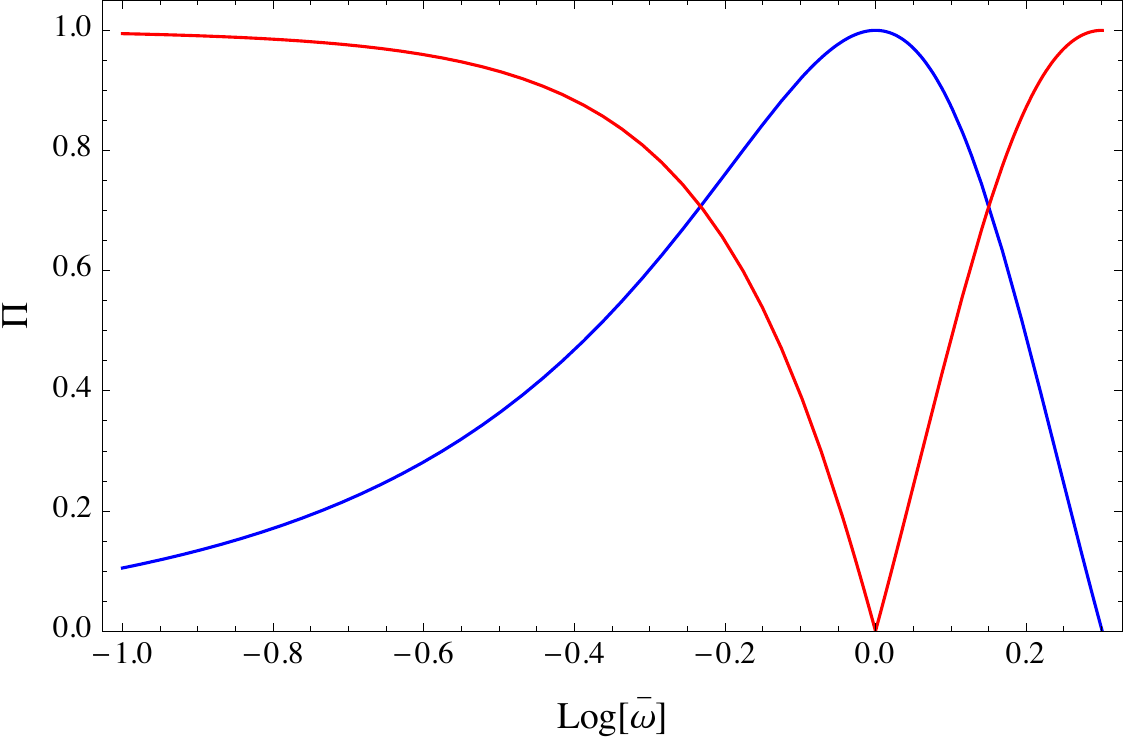}
    \includegraphics[width = 0.45\linewidth, trim = 0 0 0 0, clip]{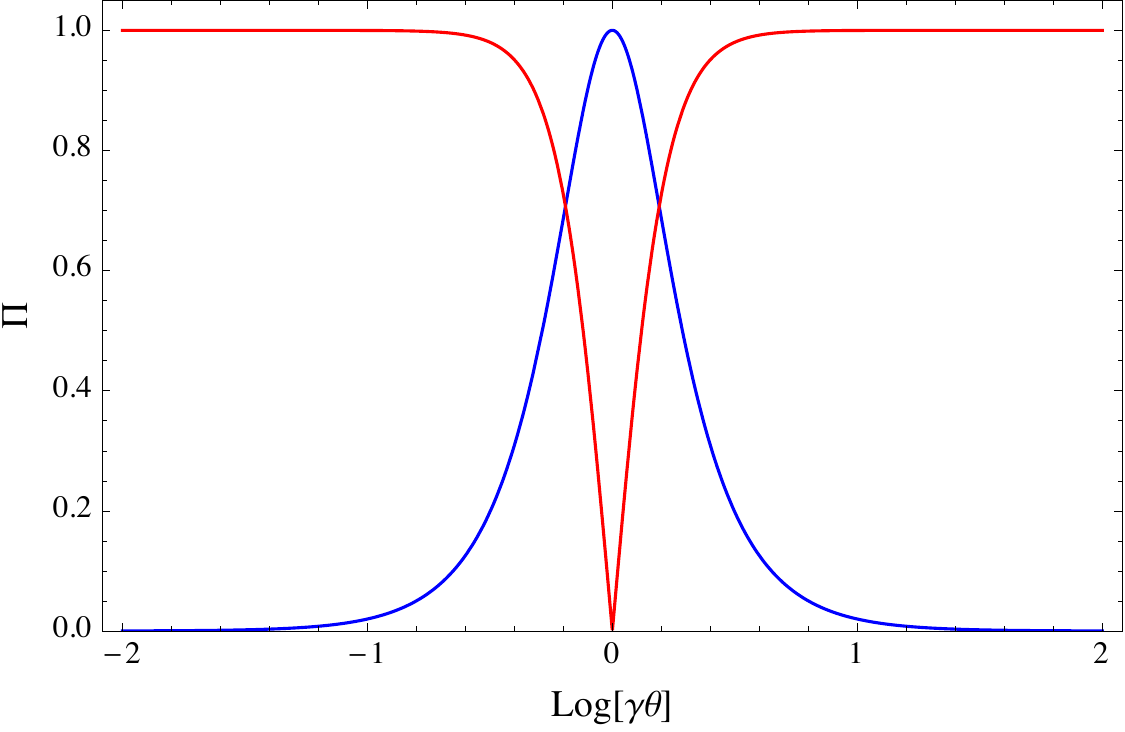}
    \caption{The relations between the polarization degree $\Pi$ and the dimensionless frequency $\hat\omega$ and the viewing angle $\gamma\theta$ for a single radiating particle with $\gamma\psi\ll1$. The blue and red lines correspond to the linear and circular polarization degrees, respectively. The top panel shows the polarization degree $\Pi$ as the function of the dimensionless frequency $\bar\omega$. The bottom panel shows the polarization degree $\Pi$ as the function of the viewing angle $\gamma\theta$. The dimensionless frequency $\bar\omega$ and the viewing angle $\gamma\theta$ are related via $\bar\omega=2/(1+\gamma^2\theta^2)$.}\label{pol_freq1} 
\end{figure}

Since the polarization degree is Lorentz invariance, according to Eq.(\ref{SPA_pil}), Eq.(\ref{SPA_piv}) and Eq.(\ref{costhe}), the linear and circular polarization degrees are
\begin{align}
&\pi_L=\left|\frac{2\gamma^2\theta^2}{1+\gamma^4\theta^4}\right|=\left|\frac{2\bar\omega-\bar\omega^2}{\bar\omega^2-2\bar\omega+2}\right|,\label{SPA_piL}\\
&\pi_V=\left|\frac{1-\gamma^4\theta^4}{1+\gamma^4\theta^4}\right|=\left|\frac{2\bar\omega-2}{\bar\omega^2-2\bar\omega+2}\right|.\label{SPA_piV}
\end{align}
In particular, $\gamma\theta>1$ and $\gamma\theta<1$ correspond to the opposite (left and right) circular polarization, respectively.
In Figure \ref{pol_freq1}, we plot the linear and circular polarization degrees $\Pi_i$ with $i=L~{\rm and}~V$ as the functions of the dimensionless frequency $\bar\omega$ and viewing angle $\gamma\theta$, respectively. The blue and red lines correspond to the linear and circular polarization degrees, respectively. The top and bottom panels show the polarization degree $\Pi$ as the functions of the dimensionless frequency $\bar\omega$ and the viewing angle $\gamma\theta$, respectively. 
We can see that high linear polarization (low circular polarization) mainly occur at $\bar\omega\sim1$ and $\gamma\theta\sim1$, otherwise, low linear polarization (high circular polarization) is dominant.

\begin{figure}
    \centering
    \includegraphics[width = 0.5\linewidth, trim = 0 0 0 0, clip]{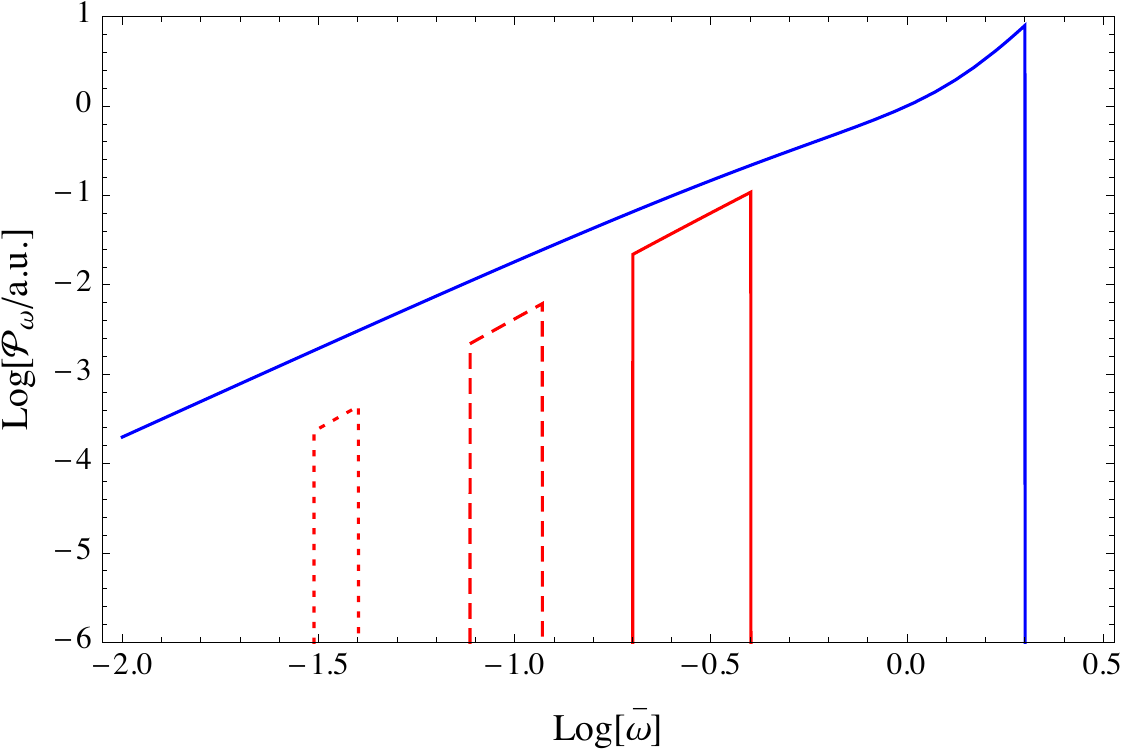}
    \caption{The spectra that are applicable for multiple radiating particles with $\gamma\psi\ll1$. The blue and red line corresponds to the on-beam case given by Eq.(\ref{syn1}) and the off-beam case given by Eq.(\ref{syn2}), respectively. The red solid, dashed and dotted lines correspond to $\gamma\Theta=3,5~{\rm and}~8$, respectively. Here $\gamma\Theta_j=1$ is taken. The unit of $\mathcal{P}_\omega$ is arbitrary. For easy comparison with different scenarios, the spectra of the off-beam cases are multiplied by an arbitrary factor in this figure.}\label{syn_spectrum} 
\end{figure}

Furthermore, we consider that multiple radiating particles are uniformly distributed in a three-dimensional beaming with an opening angle $\Theta_j$. The number of the charged particles within $(\Theta,\Theta+d\Theta)$ is
\begin{align}
N_e(\Theta)d\Theta=N_{e,0}\sin\Theta d\Theta.\label{syn_dist}
\end{align}
When the viewing direction points to the beaming center $\Theta=0$, the radiation spectrum by multiple particles might be approximately given by
\begin{align}
&\mathscr{P}_\omega(\Theta=0)\propto\int_0^{\Theta_j}\mathcal{P}_\omega(\gamma\Theta)\sin\Theta d\Theta\nonumber\\
&\propto\bar\omega^4-2\bar\omega^3+2\bar\omega^2~~~\text{for}~\bar\omega_{\min}=\frac{2}{1+\gamma^2\Theta_j^2}<\bar\omega<2,\label{syn1}
\end{align}
where $\Theta_j\ll1$ is assumed in the above equation.
Outside the particles' beam $\Theta\gg\Theta_j$, the radiation spectrum by multiple particles might be approximately given by
\begin{align}
&\mathscr{P}_\omega(\Theta)\propto\int_{\Theta-\Theta_j}^{\Theta}\mathcal{P}_\omega(\gamma\Theta)d\Theta\propto\frac{\bar\omega^5-2\bar\omega^4+2\bar\omega^3}{(2\bar\omega-\bar\omega^2)^{1/2}}\nonumber\\
&\text{for}~\bar\omega_{\min}=\frac{2}{1+\gamma^2\Theta^2}<\bar\omega<\bar\omega_{\max}=\frac{2}{1+\gamma^2(\Theta-\Theta_j)^2}.\label{syn2}
\end{align}
Notice that a factor of $\sin\Theta$ should be involved in the above equation for the viewing direction outside the particles' beam. 
In Figure \ref{syn_spectrum}, we plot the spectra for $\Theta_j=0$ and $\Theta\gg\Theta_j$ given by Eq.(\ref{syn1}) and Eq.(\ref{syn2}), respectively. For the case of $\Theta_j=0$, the peak frequency is at $\bar\omega\sim2$ near which the spectrum is narrow due to $\mathscr{P}_\omega\propto\bar\omega^4$. For the case of $\Theta\gg\Theta_j$, the spectrum is much narrower, i.e., $\Delta\bar\omega/\bar\omega\simeq2\Theta_j/\Theta\ll1$. 
Meanwhile, the larger the viewing angle $\Theta$, the narrower the spectrum and the lower the peak radiation power. At last, we should notice that the above discussion assumes that all radiating particles have the same Lorentz factor $\gamma$. The distribution of $\gamma$ for multiple particles would make the spectra relatively wider.

\begin{figure}
    \centering
    \includegraphics[width = 0.45\linewidth, trim = 0 0 0 0, clip]{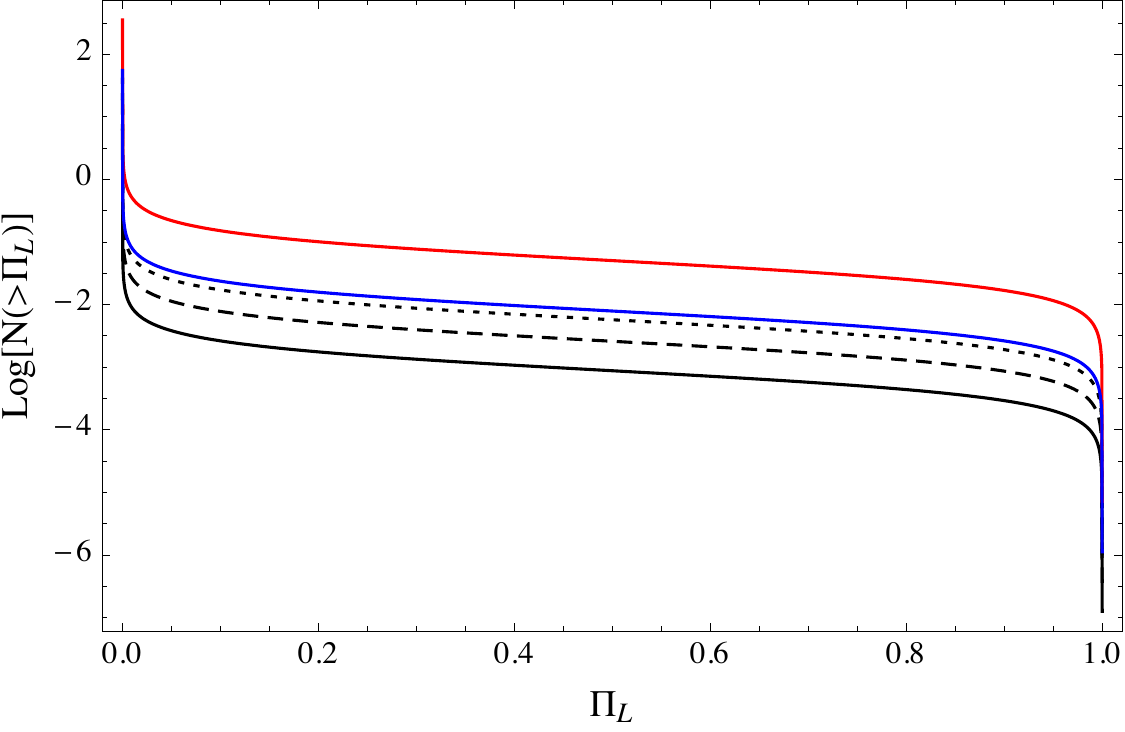}
    \includegraphics[width = 0.45\linewidth, trim = 0 0 0 0, clip]{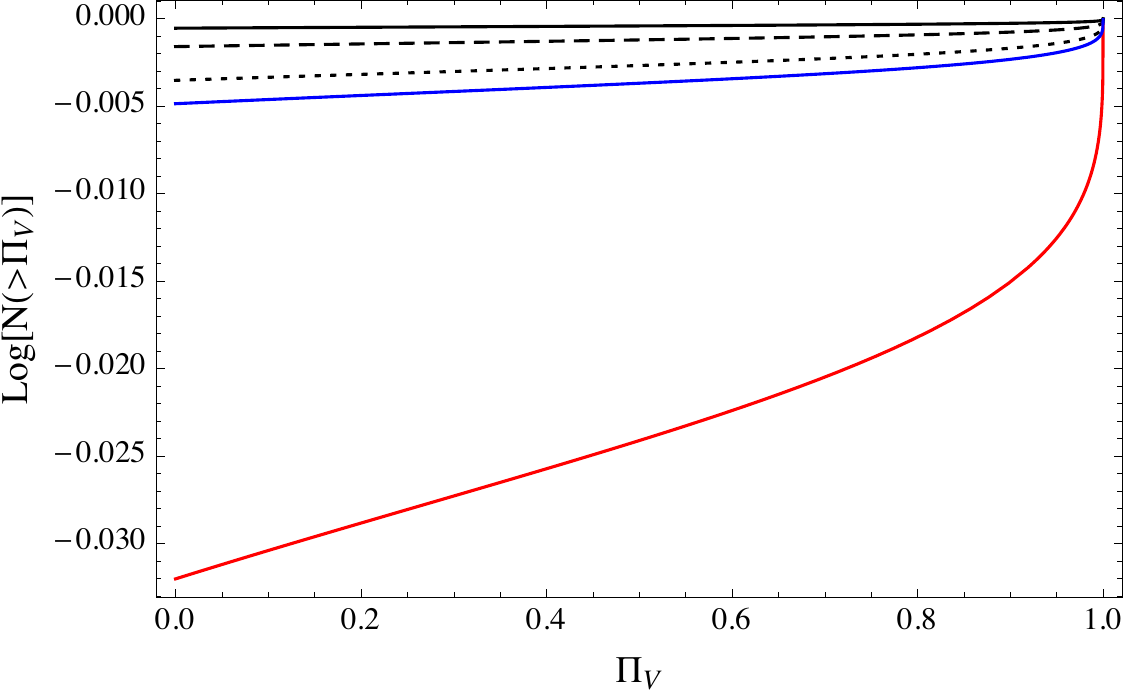}
    \caption{The cumulative distribution of the linear (top panel) and circular (bottom panel) polarization degrees. The distribution function of the viewing direction is assumed to satisfy Eq.(\ref{syn_dist}). The black solid, dashed and dotted lines correspond to $\gamma\Theta_j=3,10~{\rm and}~30$, respectively for $\gamma\theta_{\rm th}=100$. The solid red, blue and black lines correspond to $\gamma\theta_{\rm th}=10,30~{\rm and}~100$, respectively for $\gamma\Theta_j=3$. $\gamma=100$ is taken here. Different from Figure \ref{cum_dist1}, the $y$-axis is $\log[N(>\Pi_L)]$ here.}\label{cum_dist3} 
\end{figure}

The observed linear and circular polarization degrees depend on the gyration directions of the radiating particles and whether the viewing direction is within $\Theta_j$. If all radiating particles have the same gyration directions, clockwise or anticlockwise, the linear and circular polarization degrees of the multiple particles might be written as
\begin{align}
\Pi_L\simeq\left\{
\begin{aligned}
&0,&&\text{for}~\Theta\leqslant\Theta_j,\\
&\pi_L(\Theta-\Theta_j),&&\text{for}~\Theta>\Theta_j, 
\end{aligned}
\right.
\end{align}
and  
\begin{align}
\Pi_V\simeq\left\{
\begin{aligned}
&1,&&\text{for}~\Theta\leqslant\Theta_j,\\
&\pi_V(\Theta-\Theta_j),&&\text{for}~\Theta>\Theta_j, 
\end{aligned}
\right.
\end{align}
respectively, similar to the discussion in Section \ref{mechanism1}.
Notice that different from the above scenario with $\gamma\psi\gg1$, the on-beam radiation for $\gamma\psi\ll1$ is dominated by the $\sim100\%$ circular polarization.
On the other hand, if the gyration directions of the particles are random, their polarizations would cancel out. In the following discussion, we are mainly interested in the former scenario. 

Since the viewing direction is random in the solid angle, the number of the observable burst within $(\Theta,\Theta+d\Theta)$ is given by
\begin{align}
N(\Theta)d\Theta=\sin\Theta d\Theta~~~\text{for}~~~0\leqslant\Theta\leqslant\Theta_j+\theta_{\rm th},
\end{align}
otherwise, $N(\Theta)d\Theta=0$,
where $\Theta_j+\theta_{\rm th}$ corresponds to the lower limit of the frequency $\bar\omega$ or the lower limit of $\mathscr{P}_\omega$.
According the polarization properties given by Eq.(\ref{SPA_piL}) and Eq.(\ref{SPA_piV}), 
the polarization is $\sim100\%$ circular except $\gamma\theta\sim1$.
Therefore, the cumulative distribution of the linear polarization degrees for multiple radiating particles is
\begin{align}
N(>\Pi_L)\simeq\frac{\sin\Theta_j\Delta\theta_L}{1-\cos(\Theta_j+\theta_{\rm th})},
\end{align}
where $\Delta\theta_L=\theta_{L,2}-\theta_{L,1}$ with $\theta_{L,i}$ solved by Eq.(\ref{SPA_piL}), and $\Delta\theta_L\ll1$ is used in the above equation.
\begin{align}
\Delta\theta_L=\frac{1}{\gamma\Pi_L^{1/2}}\left[\left(1+\sqrt{1-\Pi_L^2}\right)^{1/2}-\left(1-\sqrt{1-\Pi_L^2}\right)^{1/2}\right].
\end{align}
Similarly, the cumulative distribution of the circular polarization degree is
\begin{align}
N(>\Pi_V)\simeq1-\frac{\sin\Theta_j\Delta\theta_V}{1-\cos(\Theta_j+\theta_{\rm th})}
\end{align}
where $\Delta\theta_V$ is solved by Eq.(\ref{SPA_piV}), leading to
\begin{align}
\Delta\theta_V=\frac{2}{\gamma}\left(\frac{1-\Pi_V}{1+\Pi_V}\right)^{1/4}.
\end{align}
In Figure \ref{cum_dist3}, we plot the cumulative distribution of the linear (top panel) and circular (bottom panel) polarization degrees, respectively. We can see that for multiple radiating particles, the polarization of this radiation mechanism is almost $100\%$ circular polarization. Thus, if the radiation mechanism of FRBs corresponds to this scenario with $\gamma\psi\ll1$, the observed moderate number fraction between linearly polarized bursts and circularly polarized bursts of FRB 20201124A \citep{Jiang22} might be due to the propagation effects, because the intrinsic polarization of this mechanism is predicted to be $\sim100\%$ circular polarized.
Notice that this result is based on the assumption that all radiating particles have the same gyration direction. If the gyration directions of the particles are random, their polarizations would cancel out.


\end{document}